\documentclass[journal]{IEEEtran}

\usepackage[linesnumbered,ruled]{algorithm2e}
\usepackage{graphicx,amssymb,mathrsfs,amsmath,pifont,amscd,latexsym,amsfonts,amsthm,colortbl}
\usepackage{graphics}
\usepackage{epstopdf}
\usepackage{float}
\usepackage{subfigure}
\usepackage{epsfig}
\usepackage{amsfonts}
\usepackage{setspace}
\usepackage{booktabs}
\usepackage{xcolor}
\usepackage{multirow}
\usepackage{soul}
\usepackage[colorlinks,urlcolor=blue,driverfallback=dvipdfm]{hyperref}

\newcommand{\norm}[1]{\lVert#1\rVert}
\usepackage{algpseudocode}
\usepackage{cite}

\DeclareMathOperator{\F}{F}
\DeclareMathOperator{\T}{T}

\begin{document}
\title{Endmember-Guided Unmixing Network (EGU-Net): A General Deep Learning Framework for Self-Supervised Hyperspectral Unmixing}

\author{Danfeng Hong,~\IEEEmembership{Senior Member,~IEEE,}
        Lianru Gao,~\IEEEmembership{Senior Member,~IEEE,}
        Jing Yao,
        Naoto Yokoya,~\IEEEmembership{Member,~IEEE,}
        Jocelyn Chanussot,~\IEEEmembership{Fellow,~IEEE,}
        Uta Heiden,~\IEEEmembership{Member,~IEEE,}
        and Bing Zhang,~\IEEEmembership{Fellow,~IEEE}

\thanks{This work was supported in part by the National Natural Science Foundation of China under Grant 42030111 and Grant 41722108, by the MIAI@Grenoble Alpes (ANR-19-P3IA-0003) and the AXA Research Fund, and by the Japan Society for the Promotion of Science (KAKENHI 18K18067). (\emph{Corresponding author: Lianru Gao})}
\thanks{D. Hong is with the Remote Sensing Technology Institute (IMF), German Aerospace Center (DLR), 82234 Wessling, Germany, and also with the Univ. Grenoble Alpes, CNRS, Grenoble INP, GIPSA-lab, 38000 Grenoble, France. (e-mail: danfeng.hong@dlr.de)}
\thanks{L. Gao and J. Yao are with the Key Laboratory of Digital Earth Science, Aerospace Information Research Institute, Chinese Academy of Sciences, Beijing 100094, China. (e-mail: jasonyao92@gmail.com; gaolr@aircas.ac.cn)}
\thanks{N. Yokoya is with Department of Complexity of Science and Engineering, The University of Tokyo and with Geoinformatics Unit, RIKEN Center for Advanced Intelligence Project (AIP), Japan. (email: yokoya@k.u-tokyo.ac.jp)}
\thanks{J. Chanussot is with the Univ. Grenoble Alpes, INRIA, CNRS, Grenoble INP, LJK, F-38000 Grenoble, France, and also with the Aerospace Information Research Institute, Chinese Academy of Sciences, Beijing 100094, China. (e-mail: jocelyn@hi.is)}
\thanks{U. Heiden is with the Remote Sensing Technology Institute (IMF), German Aerospace Center (DLR), 82234 Wessling, Germany. (e-mail: uta.heiden@dlr.de)}
\thanks{B. Zhang is with the Key Laboratory of Digital Earth Science, Aerospace Information Research Institute, Chinese Academy of Sciences, Beijing 100094, China, and also with the College of Resources and Environment, University of Chinese Academy of Sciences, Beijing 100049, China. (e-mail: zb@radi.ac.cn)}
}

\markboth{IEEE Transactions on Neural Networks and Learning Systems,~Vol.~XX, No.~XX, ~XXXX,~2021}%
{Shell \MakeLowercase{\textit{et al.}}: Endmember-Guided Unmixing Network (EGU-Net): A General Deep Learning Framework for Self-Supervised Hyperspectral Unmixing}

\maketitle
\begin{abstract}
\textcolor{blue}{This is the pre-acceptance version, to read the final version please go to IEEE Transactions on Neural Networks and Learning Systems on IEEE Xplore.} Over the past decades, enormous efforts have been made to improve the performance of linear or nonlinear mixing models for hyperspectral unmixing, yet their ability to simultaneously generalize various spectral variabilities and extract physically meaningful endmembers still remains limited due to the poor ability in data fitting and reconstruction and the sensitivity to various spectral variabilities. Inspired by the powerful learning ability of deep learning, we attempt to develop a general deep learning approach for hyperspectral unmixing, by fully considering the properties of endmembers extracted from the hyperspectral imagery, called endmember-guided unmixing network (EGU-Net). Beyond the alone autoencoder-like architecture, EGU-Net is a two-stream Siamese deep network, which learns an additional network from the pure or nearly-pure endmembers to correct the weights of another unmixing network by sharing network parameters and adding spectrally meaningful constraints (e.g., non-negativity and sum-to-one) towards a more accurate and interpretable unmixing solution. Furthermore, the resulting general framework is not only limited to pixel-wise spectral unmixing but also applicable to spatial information modeling with convolutional operators for spatial-spectral unmixing. Experimental results conducted on three different datasets with the ground-truth of abundance maps corresponding to each material demonstrate the effectiveness and superiority of the EGU-Net over state-of-the-art unmixing algorithms. The codes will be available from the website: \url{https://github.com/danfenghong/IEEE_TNNLS_EGU-Net}.
\end{abstract}

\graphicspath{{images/}}


\section{Introduction}
\IEEEPARstart{A}{irborne} or spaceborne spectroscopy (or hyperspectral) imagery is a remotely sensed three-dimensional imaging product of collecting hundreds or thousands of images finely sampled from the continuous electromagnetic spectrum, yielding very rich spectral information \cite{hong2021interpretable}. The resulting narrower swath width in hyperspectral imagery (HSI) enables the discrimination and detection of materials of interest at a more accurate level, particularly for some categories that are extremely similar each other in the range of visual light. This, therefore, has led to an increasing interest in hyperspectral data processing and analysis, such as dimensionality reduction \cite{wang2017dimensionality,hong2019learning},
image fusion and enhancement \cite{hong2019cospace,dian2019learning,hong2021more}, image segmentation and classification \cite{yang2018self,peng2018maximum,hang2019cascaded,hong2020graph}, spectral unmixing \cite{bioucas2012hyperspectral,uezato2016incorporating,thouvenin2016hyperspectral,drumetz2016blind,uezato2018hyperspectral,hong2019augmented}, and object / anomaly detection \cite{kang2017hyperspectral,xu2018joint,wu2019orsim}. Due to the low spatial resolution of HSI, there are a large number of mixed pixels in HSI, inevitably degrading spectral discrimination. 

\begin{figure*}[!t]
	  \centering
		\subfigure[linear mixing]{
			\includegraphics[width=0.315\textwidth]{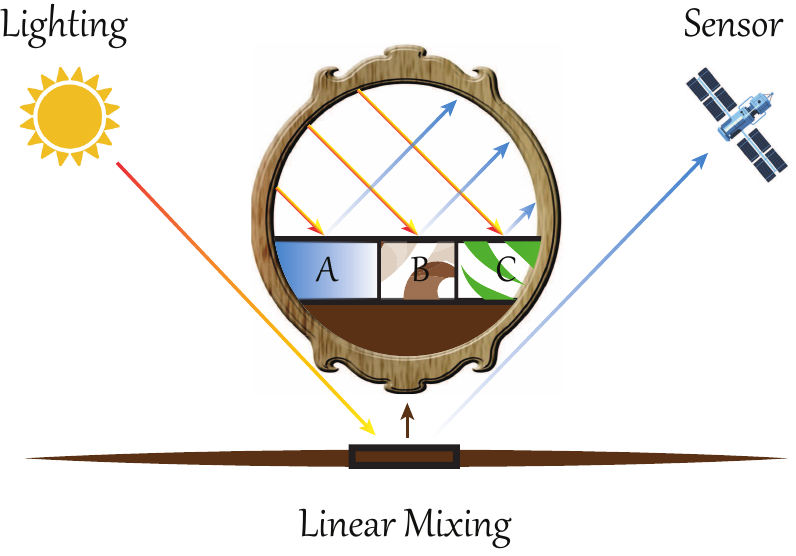}
		}
		\subfigure[nonlinear: intimate mixing]{
			\includegraphics[width=0.315\textwidth]{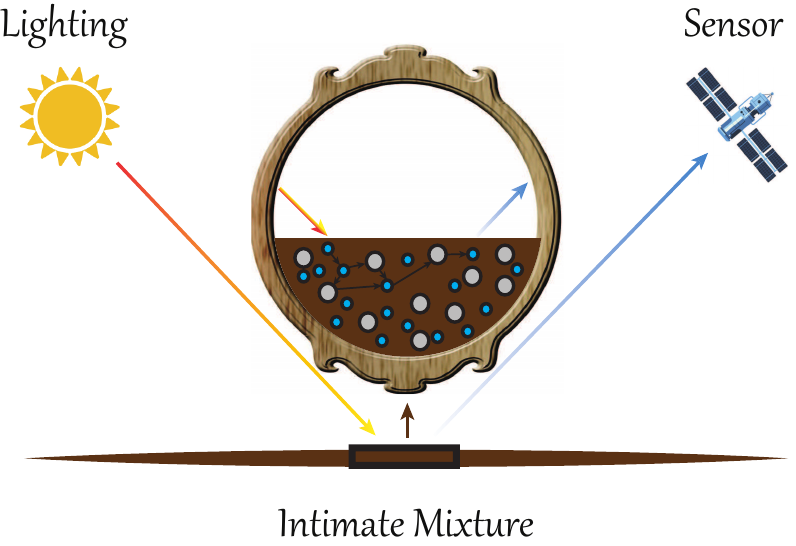}
		}
		\subfigure[nonlinear: multilayered case]{
			\includegraphics[width=0.315\textwidth]{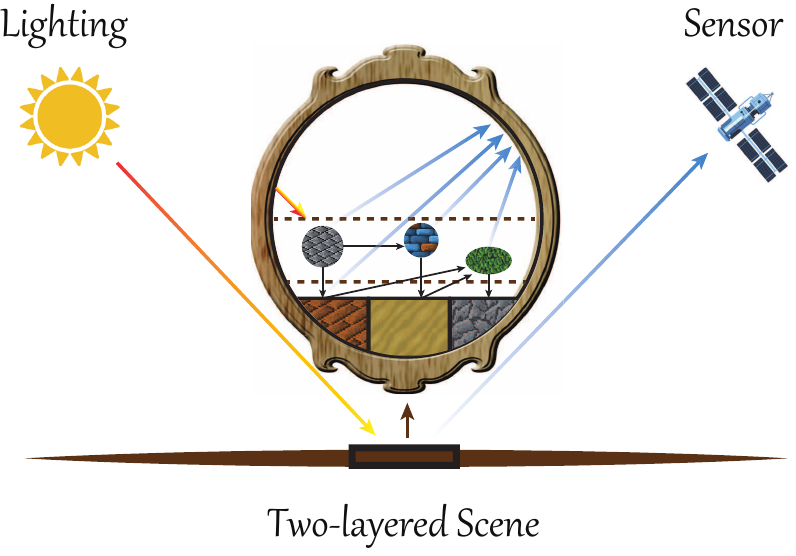}
		}
         \caption{Linear and nonlinear mixing scenarios: (a) linear mixing. (b) nonlinear mixing of intimate mixture. (c) nonlinear mixing of multilayered scattering: a two-layered case.}
\label{fig:mixing}
\end{figure*}

Spectral unmixing is usually viewed as an important pre-processing procedure prior to the high-level data analysis, aiming at simultaneously estimating a collection of individual components (termed as \textit{endmembers}) and a series of corresponding fractional percentages (termed as \textit{abundances}). In ideal conditions, that is, there are no additional errors in spectral imaging either from internal uncertainties of observed spectra or external nonlinear mixing between different materials, the unmixing process can be perfectly modeled by a linear mixing model (LMM) \cite{bioucas2012hyperspectral}, as shown in Fig. \ref{fig:mixing}(a). However, the LMM hardly makes an accurate unmixing in reality, due to the ubiquitous error inevitably existed in spectral signatures that passively transfers the unpredicted errors into LMM. In this paper, we uniformly call these general errors the spectral variability (SV). These SVs might consist of scaling factors caused by illumination and topography change, complex noise from environmental condition or instrumental sensors, physically and chemically atmospheric effects, and nonlinear mixing of materials, which, as often as not, leads to the deformation of spectral signatures and further makes it very difficult to accurately unmix these mixed pixels. Up to present, hyperspectral unmixing (HU), as a variant of blind source separation, is still a hot-spot and challenging problem in hyperspectral remote sensing.

\subsection{Motivation and Objectives}
Due to the existence of the complex SV, powerful nonlinear models are preferred. Unlike those traditional nonlinear HU approaches that might only work well in several special cases, highly powerful learning ability of deep learning (DL) techniques makes it possible to nonlinearly unmix the HSI in a more generalized way. Most previously-developed DL-based unmixing approaches basically follow autoencoder (AE)-like architecture. Despite being able to finely reconstruct the data, these methods tend to generate physically meaningless endmembers in practice, since there is a lack of effective guidance for real endmembers in the blind HU. This could further lead to poor estimation on abundances and high model's uncertainty resulted from the sensitivity to SVs.

We have to admit, however, that the DL-based strategy is a feasible and potential solution for the HU task to make a good trade-off between unmixing accuracy and generalization ability (against SVs) of the model. Therefore, this motivates us to develop a general DL-based framework accounting for the endmember information for high-performance HU.

\subsection{Method Overview and Contributions}
Towards the aforementioned goal, an endmember-guided unmixing network (EGU-Net) is proposed by fully considering the physically meaningful information of endmembers extracted from HSI. EGU-Net follows a two-stream Siamese deep network architecture for blind HU in a self-supervised fashion. One stream is to abstract the properties of endmembers obtained from the hyperspectral scene in a layer-wise manner and transfer them into another unmixing network stream via parameter sharing strategy. In addition, non-negativity and sum-to-one constraints are wisely embedded into network learning in the form of ReLU and softmax layers, respectively. More generally, the proposed EGU-Net is not only applicable to pixel-wise HU with deep neural network (DNN) but also to spatial-spectral HU with convolutional neural network (CNN). More specifically, main contributions in this paper can be summarized as follows:
\begin{itemize}
    \item Owing to the powerful learning ability of DL techniques in data representation and fitting, we heuristically develop a general deep learning framework, called EGU-Net, to address the issue of nonlinear blind HU in a more effective and generalized way.
    \item The proposed general framework consists of two different versions: one is only considering the pixel-level HU dominated by DNN, another is jointly considering spatial-spectral information to unmix the HSI with CNN-dominated architecture.
    \item Beyond the classic AE-like unmixing architecture, an end-to-end two-stream deep unmixing network is proposed to model the physically meaningful properties of real endmembers by the means of a novel self-supervised strategy. To the best of our knowledge, this is the first time to consider such endmember information in the DL-based spectral unmixing.
    \item Due to the lack of ground truth, these unmixing algorithms are hardly able to be assessed quantitatively. To this end, we provide a new hyperspectral scene over Munich with functional ground truth for HU. The ground truth can be generated by means of a simple but feasible processing chain.
\end{itemize}

The paper is organized as follows. Related work in unmixing is introduced in Section II. Section III presents the proposed network architecture. Section IV gives the experimental results and discussion on three datasets. Finally, a conclusion after presenting a brief summary is drawn in Section V.

\begin{figure*}[!t]
	  \centering
			\includegraphics[width=0.8\textwidth]{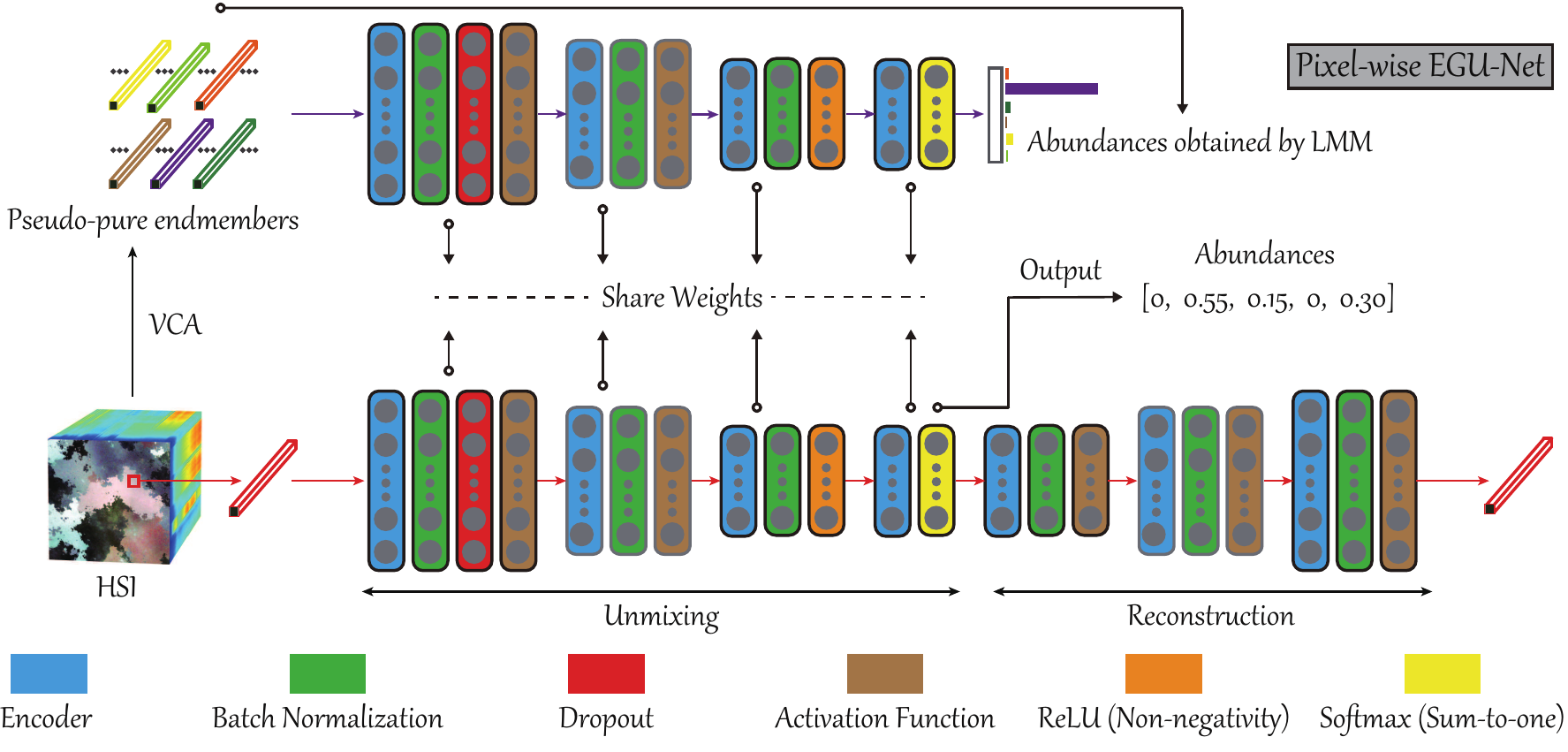}
        \caption{An overview of the proposed DL-based self-supervised unmixing framework (EGU-Net): a pixel-wise version. It consists of two subnetworks: E-Net and UR-Net, aiming at learning endmember properties and unmixing the HSI, respectively, and they can be effectively associated with shared weights in a layer-wise way. The abundances in E-Net are initialized by some existing unmixing models, i.e. LMM.}
\label{fig:EGU_Net_pw}
\end{figure*}

\section{Related Work}
\subsection{Linear Mixing Model-based Unmixing}
In the HU topics, many advanced unmixing methods on basis of LMM have been developed by an attempt to address the issue of spectral variability. Fu \textit{et al.} \cite{fu2016semiblind} developed the dictionary-adjusted unmixing approach by regrading the SV as the mismatch between endmember dictionary from the spectral library and observed spectral signatures. Similarly, authors of \cite{thouvenin2016hyperspectral} proposed the perturbed LMM (PLMM) by modeling the SV as an additive perturbation information on each endmember. One same assumption that the spectral variability should statistically follow the Gaussian distribution is made in the aforementioned two methods. In reality, there is hardly certain explicit distribution that can represent various SVs. For example, scaling factors, as a principal SV, have shown a highly coherent property with endmembers \cite{hong2017learning}. Obviously, such attributed SV can not be explained by only a perturbed error term well. For this reason, an extended LMM (ELMM) was proposed in \cite{drumetz2016blind} to model the scaling factors associated with endmembers, providing an effective solution for scaled unmixing issue from the geometrical perspective. To address other SVs that ELMM fails to consider, Hong \textit{et al.} \cite{hong2019augmented} augmented the ELMM model by learning an additional SV-based dictionary that is reasonably assumed to be incoherent or low-coherent with endmembers. The resulting augmented LMM (ALMM) shows its superiority in unmixing performance and ability to generalize various SVs. Recently, Qi \textit{et al.} \cite{qi2019region} proposed a region-based multi-view sparse regression model by incorporating a spectral library to correct the mismatches caused by SVs in HU. The same investigators in \cite{qi2019novel} further extended their work by means of a joint dictionary framework for a more effective HU. More advanced LMM-based models \cite{heylen2016hyperspectral, hong2018sulora, uezato2019hyperspectral} have sought to eliminate the effects of SVs in the process of unmixing. Nevertheless, these methods tend to meet the bottleneck particularly when unmixing the complex hyperspectral scene, due to the limitations of linearized models in data decomposition and reconstruction.

\subsection{Bilinear Mixing Model-based Unmixing}
To overcome the shortcomings of linearized unmixing techniques, some nonlinear models have been developed to enhance unmixing performance by accounting for certain particular nonlinear effects, thereby improving HU's performance to some extent. As shown in Figs \ref{fig:mixing}(b) and (c), intimate and multi-layered mixtures of materials are two kinds of common mixing behavior. One typical nonlinear model -- bilinear model (BM) in \cite{somers2009nonlinear} -- has provided a solution with an emphasis on the two-layered mixing scene. Fan \textit{et al.} \cite{fan2009comparative} improved the BM, making it applicable to more cases of material mixing. It is clear, however, that the BM-based approaches do not have the ability to generalize the HU well, as the quantity of nonlinear interactions in each pixel is only restricted in two materials. To effectively relax this constraint, a generalized bilinear model (GBM) was proposed in \cite{halimi2011nonlinear} for nonlinear unmixing. Subsequently, numerous GBM-based variants have been appeared, such as sparse and low-rank BM \cite{qu2013abundance}, GBM with semi-nonnegative matrix factorization \cite{yokoya2013nonlinear}, multi-linear mixing model \cite{heylen2015multilinear}, geometrically characterized BM \cite{yang2017nonlinear}, and adjacency effect-based BM \cite{wang2019blind}. Although these methods have gradually shown their advancement in handling the SV, particularly for the issue of multi-layered material mixture, yet they work well only in several special mixing patterns of materials due to lacking of high capability in data generalization and representation. The limitations make it difficult to cover a wider range of type of SVs.

\begin{figure*}[!t]
	  \centering
			\includegraphics[width=0.86\textwidth]{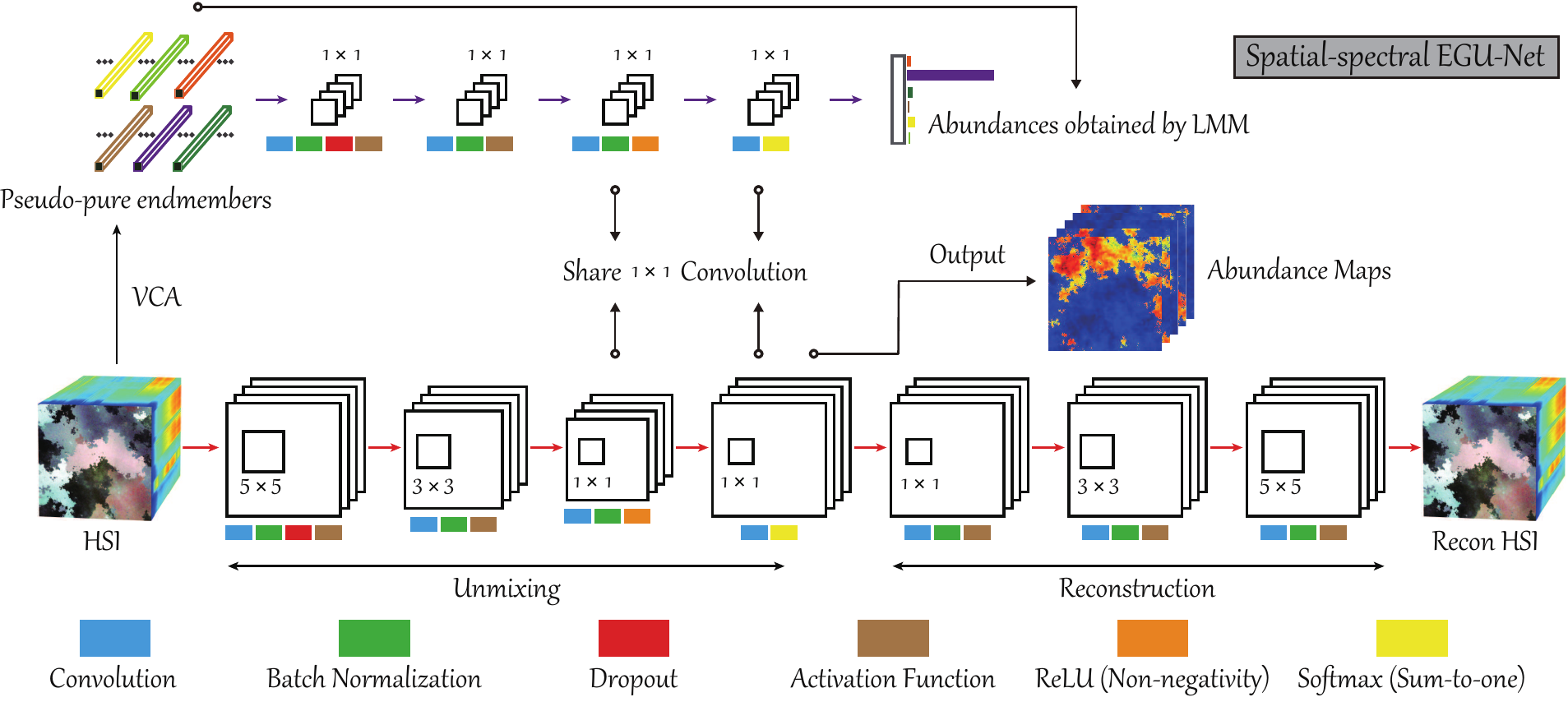}
        \caption{An overview of the proposed DL-based self-supervised unmixing framework (EGU-Net): a spatial-spectral version. The main improvement lies in the use of convolutional operations instead of fully connected encoders and the `global' unmixing with the whole HSI as the model's input. }
\label{fig:EGU_Net_ss}
\end{figure*}

\subsection{Deep Learning-based Unmixing}
The recent success of DL-based techniques has shown great power and potential in the issue of blind HU. Some exploratory efforts have been made to challenge this task with deep networks \cite{plaza2009use, plaza2009spectral,licciardi2011pixel,palsson2018hyperspectral,su2018stacked,su2019daen,wang2019nonlinear,hong2019wu}. An early work appeared in \cite{plaza2009use}, where artificial neural networks (ANNs) are used to learn the abundance fractions on a small training set. The same investigators generalized their model by intelligently selecting training samples for ANNs-based spectral mixture analysis \cite{plaza2009spectral}. Another similar work presented in \cite{licciardi2011pixel} is to unmix the spectral signatures by means of pixel-wise ANNs. These learning-based methods, as expected, perform better in spectral unmixing compared to classic linear and nonlinear models. With the recent booming development of machine learning, particularly DL, AE and its variants, as a common group of DL tool, have been widely applied to HU, achieving more competitive unmixing performance. For example, Pallson \textit{et al.} \cite{palsson2018hyperspectral} performed blind HU in the form of a neural network AE, which consists of two parts: one nonlinear unmixing and one linear reconstruction. In \cite{su2018stacked}, a two-stage AE approach that starts with a stacked AE for outlier removal and then performs HU with another AE is designed with non-negative sparse constraints for robust HU. They \cite{su2019daen} further extended their work by replacing the second-step AE with a variational AE that is capable of better inferring components and fractions from mixed materials. Similarly, authors of \cite{qu2018udas} proposed to decompose the hyperspectral data with an untied denoising sparse AE, providing a potential solution for HU. Very recently, several CNN-induced researches \cite{zhang2018hyperspectral,palsson2019convolutional} attempted to unmix the HSI in a spatial-spectral fashion, demonstrating the effectiveness of spatial information modeling in the tasks of HU. Inspired by convolutional operations in CNNs, our networks will also consider modeling spatial relationship between the target pixel and its neighbours in the process of unmixing.

\section{Endmember-Guided Unmixing Network}

In this section, we will provide a HU-targeted solution by developing a general DL framework (EGU-Net) to enhance the unmixing performance without greatly sacrificing the model's generalization ability in the process of network learning. 

The proposed EGU-Net is a self-supervised two-stream end-to-end network, including endmember network (E-Net) and unmixing-reconstruction network (UR-Net). The former aims at learning hierarchical representations of the endmembers by mapping the pseudo-pure endmembers extracted from the real HS scene to their one-hot-like abundances obtained by simple HU approaches (e.g., LMM). The latter is a common encoder-decoder architecture, which is usually taken as a backbone in unmixing networks. Significantly, parameter sharing strategy enables compact relation between the two streams, which is capable of better interacting specific information into the whole network system. Moreover, \textit{E-Net} is able to transfer the potentially intrinsic properties of the endmembers into UR-Net, and in turn the feedback from UR-Net that takes global pixels into account can also guide the E-Net towards more stable and efficient network learning. Fig. \ref{fig:EGU_Net_pw} illustrates a basic overview of the two-stream network architecture (EGU-Net).

\subsection{Endmember Network (E-Net)}

To stabilize the process of blind HU and characterize the physical significance in the estimations of endmembers and abundances, we propose to learn an additional network from the input of some relatively pure endmembers obtained from the HSI via certain endmember extraction methods, such as vertex component analysis (VCA) \cite{nascimento2005vertex}, and the corresponding output of approximate one-hot encoding abundances given by existing unmixing models (e.g., LMM). This is a typical self-supervised learning strategy without any manual labeling involved. More specifically, to provide sufficient endmember samples for training \textit{E-Net}, we extract spectral bundles\footnote{Spectral bundle is a classic and widely-used endmember extraction method to alleviate the effects of SVs effectively in the unmixing process.}~\cite{somers2012automated,drumetz2016blind} from the HSI. Inspired by \cite{somers2012automated}, the spectral bundles of endmembers used in \textit{E-Net} can be obtained by following three steps.

\begin{itemize}
    \item[1)] The HSI is divided into partially overlapped blocks, the number of blocks is determined by default given in \cite{somers2012automated};
    \item[2)] VCA is used to extract endmembers from each block, the number of endmembers is automatically estimated by HySime \cite{bioucas2008hyperspectral};
    \item[3)] A clustering algorithm, e.g., K-Means, is applied to remove repeated endmembers and aggregate all extracted endmembers to $K$ clusters. The $K$ value is experimentally and empirically set to around 20\% of all hyperspectral pixels in our cases by referring to the setting in \cite{somers2012automated}.
\end{itemize}
It should be noted that such spectral bundles are of benefit against SVs, since multiple endmembers corresponding to the same material with different scaling factors, perturbed information, and noises can be obtained in spectral bundles. This enables us to fully consider and model the SV in the training phase of \textit{E-Net}, which is conducive to more accurate abundance estimation.

Given the extracted endmembers, denoted as $\{\mathbf{x}_{i}\}_{i=1}^{N_{e}}\in \mathbb{R}^{B}$ with $B$ bands by $N_{e}$ pixels, and their nearly-pure abundances $\{\mathbf{y}_{i}\}_{i=1}^{N_{e}}\in \mathbb{R}^{C}$ with $C$ categories, the $i$-th endmember representation in the $l$-th encoder layer of \textit{E-Net}, defined by $\mathbf{z}_{i}^{(l)}$, can be then formulated as follows 
\begin{equation}
\label{eq1}
\begin{aligned}
       \mathbf{z}^{(l)}_{i}=
       \begin{cases}
       g(\mathbf{W}_{e}^{(l)}, \mathbf{b}_{e}^{(l)}, \mathbf{x}_{i}), & l=1,\\
       g(\mathbf{W}_{e}^{(l)}, \mathbf{b}_{e}^{(l)}, \mathbf{z}_{i}^{(l-1)}), & l=2,...,m,
       \end{cases}
\end{aligned}
\end{equation}
where $g(\cdot)$ is defined as the nonlinear activation function. The variables $\mathbf{W}_{e}$ and $\mathbf{b}_{e}$ represent the collections containing the weights and biases of all layers ($l=1,2,...,m$) in the \textit{E-Net}. 

Each output of encoder layer would follow up a batch normalization (BN) layer to speed up the parameter learning and alleviate the problems of exploding or vanishing gradients by reducing the internal covariance shift. Following \cite{ioffe2015batch}, the output in the BN layer is derived by
\begin{equation}
\label{eq2}
\begin{aligned}
      \mathbf{z_{BN}}^{(l)}_{i}=\gamma\mathbf{\hat{z}}_{i}^{(l)}+\beta,
\end{aligned}
\end{equation}
where $\mathbf{\hat{z}}_{i}^{(l)}$ is the $z$-score result of $\mathbf{z}_{i}^{(l)}$, $\gamma$ and $\beta$ denote the learnable network parameters.

Before feeding the $\mathbf{z_{BN}}^{(l)}_{i}$ into the activation function, a dropout layer can be employed to remove possible outliers and SVs to some extents. It should be noted, however, that in our case, the dropout layer is only activated in the first block\footnote{We define a set of layers that consists of encoder layer, BN layer, activation function layer, or dropout layer, as a block in our network architecture.}. Herein, we define $\mathbf{z_{Drop}}^{(l)}_{i}$ as the output of the dropout layer, and then the final representation ($\mathbf{a}_{i}^{(l)}$) behind an nonlinear activation in one block can be written as
\begin{equation}
\label{eq3}
\begin{aligned}
      \mathbf{a}_{i}^{(l)}=g(\mathbf{z_{Drop}}^{(l)}_{i}).
\end{aligned}
\end{equation}

As illustrated in Fig. \ref{fig:EGU_Net_pw}, we enforce the ASC and ANC in the last two blocks by means of ReLU layer and softmax layer, respectively, which are
\begin{equation}
\label{eq4}
\begin{aligned}
      \mathbf{a_{ReLU}}_{i}^{(l)}=g_{ReLU}(\mathbf{z_{BN}}^{(l)}_{i})={\rm max}(\mathbf{0}, \; \mathbf{z_{BN}}^{(l)}_{i}),
\end{aligned}
\end{equation}
and
\begin{equation}
\label{eq5}
\begin{aligned}
      \mathbf{a_{Soft}}_{i}^{(l)}=g_{Soft}(\mathbf{z}^{(l)}_{i})=\frac{e^{\mathbf{z}_{i}^{(l)}}}{\sum_{j=1}^{C}e^{\mathbf{z}_{j}^{(l)}}}.
\end{aligned}
\end{equation}
With the output of Eq. (\ref{eq5}), the cross-entropy is used to measure the loss ($L_{E}$) between $\mathbf{a_{Soft}}^{(l)}$ and $\{\mathbf{y}_{i=1}^{N_{e}}\}$ in the \textit{E-Net}
\begin{equation}
\label{eq6}
\begin{aligned}
      L_{E}=\frac{-1}{N_{e}}\sum_{i=1}^{N_{e}}[\mathbf{y}_{i}{\rm log}\mathbf{a_{Soft}}_{i}+(1-\mathbf{y}_{i}){\rm log}(1-\mathbf{a_{Soft}}_{i})].
\end{aligned}
\end{equation}

Blind SU methods usually tend to generate physically meaningless spectral signatures, e.g., corresponding to unknown or nonexistent materials, perturbed information, noises, further limiting the performance of the abundance estimation. For this reason, endmember-guided SU models have proven to be effective by means of some reference endmembers obtained by endmember extraction algorithms, e.g., VCA. In our proposed EGU-Net, \textit{E-Net} plays a similar role, i.e., to guide the unmixing process in \textit{UR-Net} by providing endmember-related properties or information (e.g., curve shape, spectral absorption in particular wavelengths reflecting material characteristics), thereby yielding better abundance estimation results.

What is more noteworthy is that unlike existing AE-based unmixing networks \cite{palsson2019convolutional,wang2019nonlinear,zhao2019hyperspectral} that directly use VCA results to initialize or regularize the decoder part to embed the endmember information, our \textit{E-Net} prefers to play the role of guidance or correction in the AE-like \textit{UR-Net} by hierarchically abstracting endmember properties and progressively transferring them into \textit{UR-Net} in a layer-by-layer manner. Plus, the nonlinear reconstruction in the decoder part in \textit{UR-Net} is more apt to yield more accurate estimation of abundance maps.

\subsection{Unmixing-Reconstruction Network (UR-Net)}

Inspired by the recent success of AE-based unmixing framework, we design a similar \textit{UR-Net}, as shown in the bottom of Fig. \ref{fig:EGU_Net_pw}, involving two same basic modules: unmixing and reconstruction. In our UR-Net, two distinctive settings might lead to effective performance improvement in unmixing applications, as listed in the following.

\begin{itemize}
\item We still adopt nonlinear activation functions in reconstruction part rather than widely-used linear ones. Due to the complexity of various SVs, the linearized operation fails to finely reconstruct the original spectral signatures. Consequently, the resulting reconstruction gap may further have a negative feedback on the to-be-estimated abundances. In other words, we hardly seek out only several linearly-reconstructed endmembers that are able to address various SVs well.
\item More importantly, the network parameters in our \textit{UR-Net} stream are partially shared with those of the \textit{E-Net} stream, making it effective to hierarchically transfer different endmember information into \textit{UR-Net}. Since the proposed EGU-Net is an end-to-end network, its subnetworks: \textit{E-Net} and \textit{UR-Net} are then jointly trained in a partial parameter sharing fashion (see Figs. \ref{fig:EGU_Net_pw} and \ref{fig:EGU_Net_ss}). Intuitively, if we denote $f_{\Phi}$ as the to-be-estimated unmixing function with respect to a given HSI, the $f_{\Phi}$ is then applicable to both the endmembers extracted from the HSI and the rest mixed pixels. Accordingly, it is natural to perform the parameter sharing strategy between \textit{E-Net} and \textit{UR-Net}.
\end{itemize}
By doing so, our network can to a great extent ensure stable and desirable unmixing results. Accordingly, the optimization problem for the \textit{UR-Net} stream is given by minimizing the following reconstruction loss ($L_{UR}$).
\begin{equation}
\label{eq7}
\begin{aligned}
     \mathop{\min}_{\mathbf{W}_{ur},\mathbf{b}_{ur}}&\norm{\mathbf{X}-f_{r}(g_{u}(\mathbf{W}_{ur},\mathbf{b}_{ur},\mathbf{X}))}_{\F}^{2},\\
     &{\rm s.t.}\;\; \mathbf{W}_{ur}=\mathbf{W}_{e}, \; \mathbf{b}_{ur}=\mathbf{b}_{e},
\end{aligned}
\end{equation}
where $f_{r}$ and $g_{u}$ correspond to the mapping functions of unmixing and reconstruction modules with respect to the weights $\mathbf{W}_{ur}$ and biases $\mathbf{b}_{ur}$ of the \textit{UR-Net}, respectively. With the trained network, the final abundances can be directly inferred via learned parameters, that is, $\mathbf{a_{Soft}}$ (see the output in Fig. \ref{fig:EGU_Net_pw}). Accordingly, the overall loss of EGU-Net is
\begin{equation}
\label{eq8}
\begin{aligned}
       L_{O}=L_{E}+L_{UR}.
\end{aligned}
\end{equation}

\subsection{Spatial-Spectral Unmixing with CNNs: An Extension}

The proposed EGU-Net, which takes the Siamese network as a backborne, is a general but effective DL framework for HU. Intuitively, it can be extended to a spatial-spectral unmixing version by means of convolutional neurons. For this reason, we attempt to model the spatial relation of the target pixel and its neighbors with CNNs in the process of unmixing. More specifically, the resulting spatial-spectral EGU-Net, EGU-Net-ss for short, holds the basically identical network architecture with pixel-wise EGU-Net (see Fig. \ref{fig:EGU_Net_pw}), EGU-Net-pw for short. The main differences between the two lie in two aspects. In one, despite same input in the \textit{E-Net}, EGU-Net-ss takes advantage of $1\times 1$ convolution kernel instead of the fully connected encoder and only shares such convolutional kernels in the last two blocks with \textit{UR-Net}. In the other, beyond the pixel-by-pixel reconstruction of spectral signatures, the \textit{UR-Net} of EGU-Net-ss is able to feed the whole HSI into the network, yielding a `global' unmixing, by paying equivalent attentions on spatial information with different-size receptive fields. Fig. \ref{fig:EGU_Net_ss} gives more details regarding EGU-Net-ss in the network architecture setting.

\begin{table}[!t]
\centering
\caption{Network configuration in each layer of our EGU-Nets: pixel-wise EGU-Net (EGU-Net-pw) and spatial-spectral EGU-Net (EGU-Net-ss). FC, Conv, and AvgPool are abbreviations of fully connected, convolution, and average pooling, respectively. The symbols of `$\leftrightarrow$' and `--' represent the parameter sharing and no operations, respectively.}
\resizebox{0.47\textwidth}{!}{ 
\begin{tabular}{c||ccc|ccc}
\toprule[1.5pt]
& \multicolumn{3}{c|}{EGU-Net-pw} & \multicolumn{3}{c}{EGU-Net-ss}\\
\hline Pathway & \textit{E-Net} & & \textit{UR-Net} & \textit{E-Net} & & \textit{UR-Net}\\
\hline \multirow{5}{*}{Block1} & FC Encoder & \multirow{5}{*}{$\leftrightarrow$} & FC Encoder & $1\times 1$ Conv  & \multirow{5}{*}{--} & $5\times 5$ Conv \\
& BN & & BN & BN & & BN\\
& Dropout & & Dropout & Dropout & & Dropout\\
& -- & & -- & -- & & $2\times 2$ AvgPool\\
& Tanh & & Tanh & Tanh & & Tanh\\
\hline \multirow{4}{*}{Block2} & FC Encoder & \multirow{4}{*}{$\leftrightarrow$} & FC Encoder & $1\times 1$ Conv & \multirow{4}{*}{--} & $3\times 3$ Conv \\
& BN & & BN & BN & & BN\\
& -- & & -- & -- & & $2\times 2$ AvgPool\\
& Tanh & & Tanh & Tanh & & Tanh\\
\hline \multirow{4}{*}{Block3} & FC Encoder & \multirow{4}{*}{$\leftrightarrow$} & FC Encoder & $1\times 1$ Conv & \multirow{4}{*}{$\leftrightarrow$} & $1\times 1$ Conv \\
& BN & & BN & BN & & BN\\
& -- & & -- & -- & & $2\times 2$ AvgPool\\
& ReLU & & ReLU & ReLU & & ReLU\\
\hline \multirow{2}{*}{Block4} & FC Encoder & \multirow{2}{*}{$\leftrightarrow$} & FC Encoder & $1\times 1$ Conv & \multirow{2}{*}{$\leftrightarrow$} & $1\times 1$ DeConv \\
& Softmax & & Softmax & Softmax & & Softmax\\
\hline \multirow{3}{*}{Block5} & -- & \multirow{3}{*}{--} & FC Decoder & -- & \multirow{3}{*}{--} & $1\times 1$ DeConv \\
& -- & & BN & -- & & BN\\
& -- & & Sigmoid & -- & & Sigmoid\\
\hline \multirow{3}{*}{Block6} & -- & \multirow{3}{*}{--} & FC Decoder & -- & \multirow{3}{*}{--} & $1\times 1$ DeConv \\
& -- & & BN & -- & & BN\\
& -- & & Sigmoid & -- & & Sigmoid\\
\hline \multirow{3}{*}{Block7} & -- & \multirow{3}{*}{--} & FC Decoder & -- & \multirow{3}{*}{--} & $3\times 3$ DeConv \\
& -- & & BN & -- & & BN\\
& -- & & Sigmoid & -- & & Sigmoid\\
\hline \multirow{3}{*}{Block8} & -- & \multirow{3}{*}{--} & FC Decoder & -- & \multirow{3}{*}{--} & $5\times 5$ DeConv \\
& -- &  & BN & -- & & BN\\
& -- &  & Sigmoid & -- & & Sigmoid\\
\bottomrule[1.5pt]
\end{tabular}}
\label{tab:network_configuration}
\end{table}

\subsection{Clarifying Details on Our Network Architecture} 
The requires to unmixing applications and effective network training drive us to particularize the network architecture in a layer-by-layer way. EGU-Net successively starts with one encoder layer, one BN layer, one dropout layer, and one activation function layer in the first block, where the dropout layer added in the beginning is to remove outliers against SVs to some extent. A ReLU layer is then specified to meet an ANC demand in the penultimate layer of the whole network, while the ASC on the abundances is added in the network by means of softmax layer. More specific network configuration in each layer is detailed in Table \ref{tab:network_configuration}.

\begin{figure}[!t]
	  \centering
		\subfigure[Synthetic Data]{
			\includegraphics[width=0.14\textwidth]{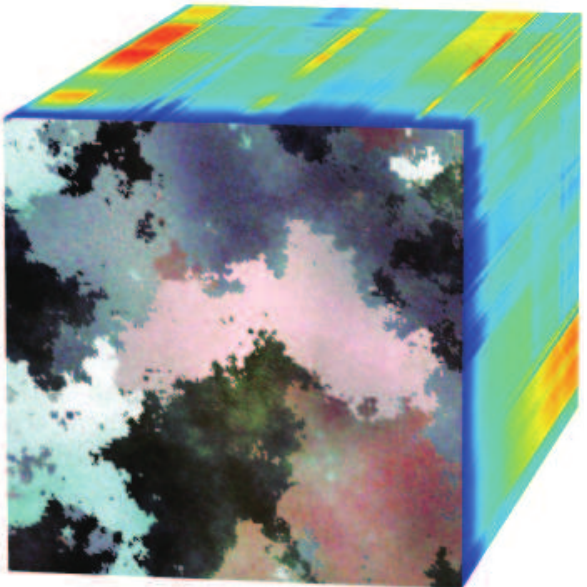}
			\label{fig:FC_simulated}
		}
		\subfigure[AVIRIS Data]{
			\includegraphics[width=0.14\textwidth]{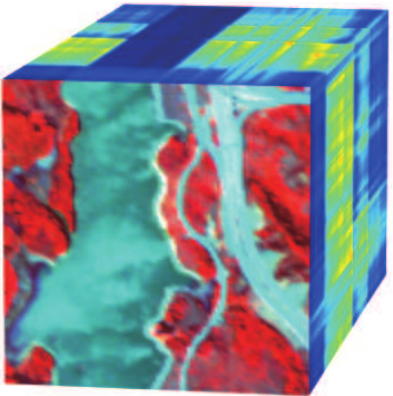}
			\label{fig:FC_river}
		}
		\subfigure[EnMAP Data]{
			\includegraphics[width=0.14\textwidth]{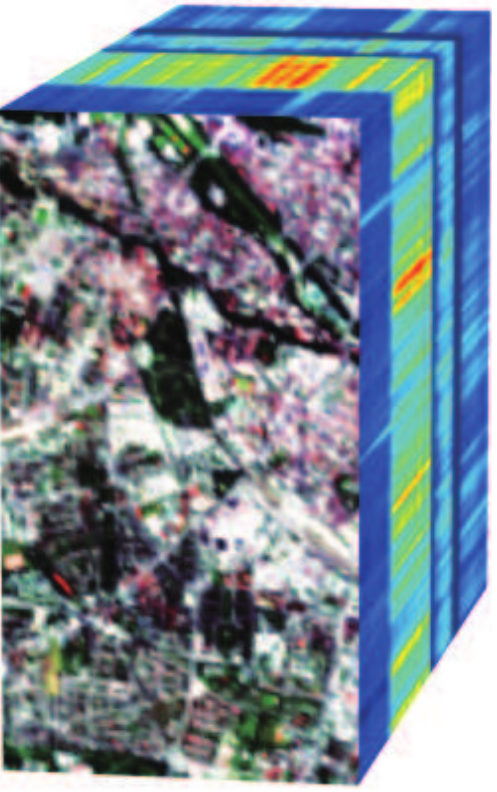}
			\label{fig:FC_munich}
		}
		\caption{Three hyperspectral datasets used for HU: (a) synthetic data, (b) AVIRIS data over Jasper Ridge, and (c) EnMAP data over Munich.}
\label{fig:FC_image}
\end{figure}

\section{Experiments}
\subsection{Data Description}

\subsubsection{Synthetic Dataset}
This dataset was simulated by selecting five reference endmembers from the United States Geological Survey (USGS) spectral library. The full image consists of 200$\times$200 pixels with 224 spectral bands covering the wavelength range from 0.4 $\mu m$ to 2.5 $\mu m$. There are various SVs in the synthetic dataset, including the principal scaling factors caused by illumination and topology change, and non-Gaussian noises as other unknown and complex SVs. More details about the simulating process (e.g., using physical-inspired Hapke model) can be found in \cite{drumetz2016blind,yao2021sparsity}. Fig. \ref{fig:FC_simulated} shows the false color cube of the HSI.

\subsubsection{Jasper Ridge Dataset}
The real hyperspectral scene was acquired by the airborne visible / infrared imaging spectrometer (AVIRIS) sensor over a rural area at Jasper Ridge, California, USA. The original image contains $512\times 614$ pixels with $224$ wavelength bands ranging from 0.38 $\mu m$ to 2.5 $\mu m$ at a ground sampling distance (GSD) of 20 $m$. We selected a widely-used region of interest (ROI) with the size of $100\times 100$ pixels and retained $198$ bands after removing noisy and water absorption bands. An example for the hyperspectral region is given in Fig. \ref{fig:FC_river}. In this studied scene, four main endmembers are investigated, i.e., \textit{\#1 Tree}, \textit{\#2 Water}, \textit{\#3 Soil}, and \textit{\#4 Road}, with referential abundance maps given from the website\footnote{\url{https://rslab.ut.ac.ir/data}}.

\subsubsection{EnMAP Munich Dataset}
The second EnMAP dataset is generated from corresponding HyMap data over an urban area, Munich, Germany, using an end-to-end EnMAP simulation tool: EeteS \cite{segl2012eetes}. We selected a ROI that comprises $93\times 171$ pixels at 30 $m$ GSD with 221 bands in the spectral range from 0.38 $\mu m$ to 2.5 $\mu m$ (see Fig. \ref{fig:FC_munich}). Five dominated materials in this scene are investigated, such as \textit{\#1 Roof}, \textit{\#2 Asphalt}, \textit{\#3 Soil}, \textit{\#4 Water}, and \textit{\#5 Vegetation}.

Besides, we will herein introduce a simple but feasible processing chain for the ground truth generation of abundances and endmmebers of HSIs by relying on a high-quality and high-resolution (e.g., HyMap's GSD) ground truth used for land cover and land use mapping. More specifically, a comprehensive processing flow is given as follows.

\begin{figure}[!t]
	  \centering
		\subfigure{
			\includegraphics[width=0.45\textwidth]{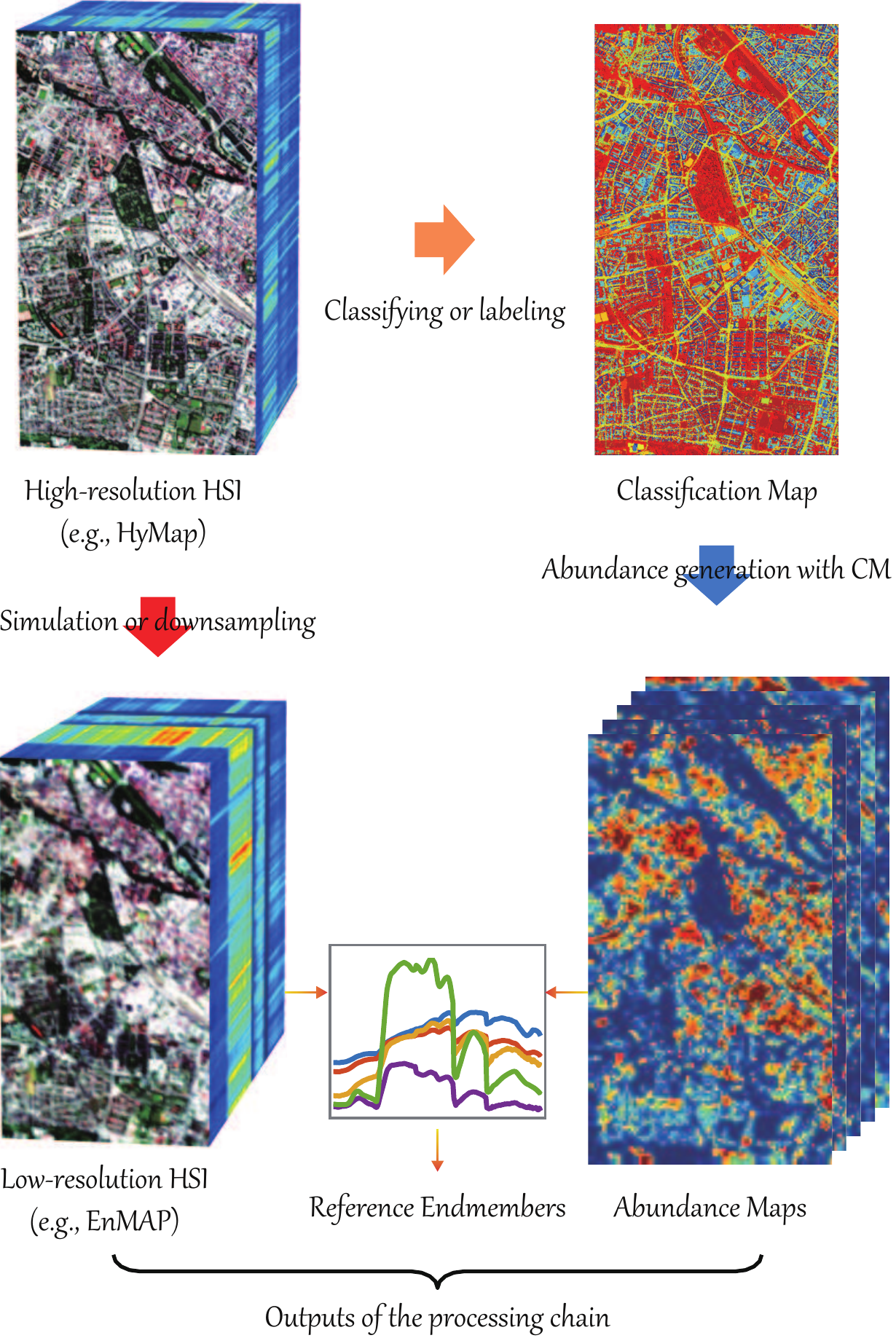}
		}
        \caption{An illustration of the processing chain for generating abundance maps and endmembers.}
\label{fig:ProcessingChain}
\end{figure}

\begin{itemize}
    \item [1)] Given high-resolution HSIs (e.g., HyMap), low-resolution mixed HSIs (e.g., EnMAP) can be obtained by using EeteS simulator or simple Gaussian downsampling.
    \item [2)] High-accuracy classification maps are generated by powerful classifiers or given by manually labeling on high-resolution HSIs.
    \item [3)] Ground truth for abundances can be computed by the classification maps and the known sampling rate (e.g., GSD) between high-resolution and low-resolution HSIs.
    \item [4)] With the generated abundances, potential endmembers can be simply selected from those pixels that only belong to one material and then the final $C$ reference endmembers can be further obtained by averaging these pure endmembers belonging to the same category. 
\end{itemize}
Fig. \ref{fig:ProcessingChain} illustrates the above workflow. The design of the processing chain just aims to provide a feasible idea and procedure to generate a convincing ground truth (e.g., abundance maps and endmembers) for the HU task. As the imaging techniques mature or new hyperspectral satellites launch successfully, the low-resolution and high-resolution images corresponding to the same scene are available easier. At that moment, we believe that the proposed processing chain would be of great help and benefit to the quantitative performance assessment of real hyperspectral scenes in the unmixing task.

\subsection{Experimental Setup}

\subsubsection{Implementation details}
Our approach is implemented on the Tensorflow framework. The model is trained on the training set, and the hyper-parameters are determined using a grid search on the validation set. In the training phase, we adopt the Adam optimizer with the ``poly'' learning rate policy. The current learning rate can be updated by multiplying the base one with $(1-\frac{iter}{maxIter})^{power}$, where the base learning rate and power are set to 0.1 and 0.99, respectively. Moreover, the momentum is parameterized by 0.9 and the rate of dropout is 0.9 in our case. The model training ends up with 200 epochs with a minibatch size same with the number of extracted endmembers used for training. The setting with regard to such a minibatch training can be specifically explained as follows. We firstly have to clarify that the unmixing problem is usually seen as regression rather than classification. That is, unlike the classification that only needs to find the position corresponding to the maximum response or score (e.g., in the softmax layer), there are higher requirements for the regression accuracy in the unmixing task. Therefore, a larger batch or full batch in \textit{E-Net} tends to step towards optimal solutions of networks accurately and stably. Furthermore, as \textit{E-Net} and \textit{UR-Net} are jointly trained in our end-to-end EGU-Net and the number of samples in UR-Net is larger than that in E-Net, we adopt the number of extracted endmembers as the minibatch size in the training phase of EGU-Net. Note that the number of these pseudo-endmembers extracted by VCA is determined by the strategy of spectral bundles presented in \cite{somers2012automated}.

\subsubsection{Evaluation metrics}
To quantitatively evaluate the unmixing performance, the abundance overall root mean square error (aRMSE) is calculated by collecting all pixel-wise abundances, as defined by 
\begin{equation}
\label{eq9}
\begin{aligned}
       aRMSE=\frac{1}{N}\sum_{i=1}^N\sqrt{\frac{1}{C}\sum_{j=1}^C(\mathbf{y}_{ji}-\hat{\mathbf{y}}_{ji})^2},
\end{aligned}
\end{equation}
and the spectral angle distance (SAD) is used to measure the similarity between the estimated endmember and the real one of each material by the following definition
\begin{equation}
\label{eq10}
\begin{aligned}
       SAD_{j}=arccos(\frac{\mathbf{e}_{j}^{\T}\mathbf{\hat{e}}_{j}}{\norm{\mathbf{e}_{j}}\norm{\mathbf{\hat{e}}_{j}}}).
\end{aligned}
\end{equation}
With the results of Eq. (\ref{eq10}), an average SAD (aSAD) index is also given by
\begin{equation}
\label{eq11}
\begin{aligned}
       aSAD=\frac{1}{C}\sum_{j=1}^{C}SAD_{j}.
\end{aligned}
\end{equation}
The parameter $C$ denotes the number of endmembers, which is $5$, $4$, and $5$ for three used hyperspectral data, respectively.

Notably, different from linear-mixture/nonlinear-fluctuation models presented in \cite{chen2012nonlinear,zhao2019hyperspectral} that can represent explicit mappings or weights (i.e., endmembers), the reconstruction (i.e., decoder part) in \textit{UR-Net} is fully nonlinear, leading to indirect endmember extraction. Alternatively, the endmembers of our model ($\mathbf{E}=\{\mathbf{e}_{j}\}_{j=1}^{C}$) can be estimated by solving a simple linear system when the abundances are given, we then have
\begin{equation}
\label{eq12}
\begin{aligned}
     \mathop{\min}_{\mathbf{E}}\norm{\mathbf{X}-\mathbf{E}\mathbf{Y}}_{\F}^{2},\;\;{\rm s.t.}\;\; \mathbf{E}\succeq \mathbf{0}.
\end{aligned}
\end{equation}
We have to clarify, however, that the use of a linear model in Eq. (\ref{eq12}) only aims to visualize the endmember spectrum and further to evaluate the quality of estimated abundances by comparing the spectral profile and absorption in certain wavelengths of endmembers with reference endmembers.

\subsection{Ablation Study on the Synthetic Data}

Linearly-reconstructed AE \cite{palsson2018hyperspectral} (DAEU) and CNN \cite{palsson2019convolutional} (CNNAEU) can be taken as the baselines for pixel-wise and spatial-spectral unmixing, respectively. There are two different modules in our network, namely \textit{E-Net} and nonlinear reconstruction part. As a result, we perform the ablation analysis on the synthetic data to investigate the effectiveness of these components for HU applications. 

Table \ref{table:Ablation} lists the performance gain in terms of aRMSE by integrating different modules, where the averaged aRMSEs are reported out of 10 runs due to the randomized network initialization and slightly different endmembers extracted by VCA (if needed). In detail, the unmixing performance of two baselines (DAEU and CNNAEU) with the linearized reconstruction is relatively inferior to that of their corresponding nonlinear versions (\textit{UR-Net}-pw and \textit{UR-Net}-ss). Unlike these fully unsupervised methods that fail to model the endmember information effectively, the self-supervised \textit{E-Net} that only considers the limited number of pseudo-endmembers has shown a desirable result at an increase of approximate 0.0040 over \textit{UR-Net}-pw in terms of aRMSE. More significantly, our proposed EGU-Nets -- EGU-Net-pw and EGU-Net-ss -- outperform the others dramatically, bringing at least 0.01 performance improvement in aRMSE compared to the baselines. This greatly demonstrates the superiority of the proposed general DL framework (EGU-Net) for the HU task.

\begin{table}
\centering
\caption{Ablation analysis on the synthetic dataset, where `linear' and `nonlinear' stand for the reconstruction types in the AE-based networks, and \textit{E-Net} represents the endmember-based additional network, while `FC' and `Conv' denote the fully connected encoder and convolutional operations, respectively. The best results are shown in bold.}
\resizebox{0.48\textwidth}{!}{ 
\begin{tabular}{l|ccccc|c}
\toprule[1.5pt]
Model & Linear & Nonlinear & \textit{E-Net} & FC & Conv & aRMSE\\
\hline\hline
DAEU & $\checkmark$ & $\times$ & $\times$ & $\checkmark$ & $\times$ & 0.0327\\
CNNAEU & $\checkmark$ & $\times$ & $\times$ & $\times$ & $\checkmark$ & 0.0249\\
\textit{UR-Net}-pw & $\times$ & $\checkmark$ & $\times$ & $\checkmark$ & $\times$ & 0.0257\\
\textit{UR-Net}-ss & $\times$ & $\checkmark$ & $\times$ & $\times$ & $\checkmark$ & 0.0232\\
\textit{E-Net} & $\times$ & $\checkmark$ & $\checkmark$ & $\checkmark$ & $\times$ & 0.0216\\
\hline
EGU-Net-pw & $\times$ & $\checkmark$ & $\checkmark$ & $\checkmark$ & $\times$ & 0.0200\\
EGU-Net-ss & $\times$ & $\checkmark$ & $\checkmark$ & $\times$ & $\checkmark$ & \bf 0.0183\\
\bottomrule[1.5pt]
\end{tabular}}
\label{table:Ablation}
\end{table}

\subsection{Unmixing Performance Comparison with State-of-the-Art}

Several state-of-the-art competitors related to the blind HU task are selected for performance comparison, including 

\noindent \textbf{Non-DL-based unmixing approaches:} fully constrained least squares unmixing (FCLSU) \cite{heinz2001fully}, partial constrained least squares unmixing (PCLSU) \cite{heylen2011fully}, sparse unmixing by variable splitting and augmented Lagrangian (SUnSAL) \cite{bioucas2010alternating}, subspace unmixing with low-rank attribute embedding (SULoRA)\footnote{\url{https://github.com/danfenghong/IEEE_JSTSP_SULoRA}} \cite{hong2018sulora}, and ALMM\footnote{\url{https://github.com/danfenghong/ALMM_TIP}}  \cite{hong2019augmented}. Note that we simply extended some of the above methods to the blind HU versions by simultaneously updating the endmembers for a fair comparison;

\noindent \textbf{DL-based unmixing approaches:} DAEU \cite{palsson2018hyperspectral}, deep autoencoder networks (DAEN) \cite{su2019daen}, CNNAEU \cite{palsson2019convolutional}, and our two versions: EGU-Net-pw and EGU-Net-ss.

Regarding the parameter setting of these existing models, default parameters given in original references are applied in our experiments for simplicity.

Table \ref{table:unmixing_simulated} lists the quantitative performance comparison in terms of aRMSE, aSAD, and SAD for each endmember on the synthetic data. Overall, FCLSU yields poor unmixing results for both abundance estimation and endmember extraction, due to the effects of various SVs. Unlike FCLSU, PCLSU assumes that the estimated abundances should be in a cone rather than in a simplex by dropping the ASC, bringing around 2.5\% performance improvement in terms of aRMSE over FCLSU. By making use of sparsity prior on the abundances effectively, SUnSAL can further enhance the unmixing performance in spite of being improved by only 0.3\%. SULoRA attempts to alleviate the effects of high-dimensional SVs by searching a low-rank subspace where the unmixing is proceeded more robustly, yielding an obvious improvement on aRMSE (about 1.4\%) and aSAD (0.06\%) as compared to the SUnSAL. Similarly, ALMM models SVs more finely, which allows a better result in aRMSE and aSAD as well as SAD for almost all endmembers.

In general, DL-based models might pose a greater potential to blindly unmix the hyperspectral data. Despite being slight lower than the two powerful linear models (SULoRA and ALMM) in terms of aRMSE, yet they, such as DAEU, DAEN, and CNNAEU, hold advantage in extracting endmembers. Not unexpectedly, the proposed simple but effective DL framework either EGU-Net-pw or EGU-Net-ss outperforms other unmixing algorithms dramatically, which indicates that our networks are capable of improving the unmixing accuracies for both abundances and endmembers in the presence of complex SVs. In all indices, the two EGU-Nets basically achieve the best estimations, in which the CNN-induced spatial-spectral unmixing more tends to generate the optimal results.

\begin{table*}
\centering
\caption{Quantitative results on the synthetic dataset, where aRMSE and aSAD as well as SAD for each material are reported with their standard deviations from 10 Monte Carlo runs. The best one is shown in bold.}
\resizebox{0.93\textwidth}{!}{ 
\begin{tabular}{l|c|ccccc|c}
\toprule[1.5pt]
Methods & aRMSE & SAD\_$\mathbf{e}_{1}$ & SAD\_$\mathbf{e}_{2}$ & SAD\_$\mathbf{e}_{3}$ & SAD\_$\mathbf{e}_{4}$ & SAD\_$\mathbf{e}_{5}$ & aSAD \\
\hline\hline
FCLSU & 6.78 $\pm$ 0.18\% & 2.75 $\pm$ 0.44\% & 2.69 $\pm$ 0.54\% & 0.65 $\pm$ 0.04\% & 1.26 $\pm$ 0.21\% & 0.70 $\pm$ 0.03\% & 1.61 $\pm$ 0.20\%\\
PCLSU & 4.33 $\pm$ 0.42\% & 0.64 $\pm$ 0.63\% & 0.82 $\pm$ 0.35\% & 0.23 $\pm$ 0.03\% & 0.57 $\pm$ 0.40\% & 0.44 $\pm$ 0.01\% & 0.54 $\pm$ 0.16\%\\
SUnSAL & 4.05 $\pm$ 0.19\% & 0.77 $\pm$ 0.55\% & 1.12 $\pm$ 0.27\% & 0.25 $\pm$ 0.05\% & 0.57 $\pm$ 0.39\% & 0.44 $\pm$ 0.01\% & 0.63 $\pm$ 0.12\%\\
SULoRA & 2.61 $\pm$ 0.21\% & 0.69 $\pm$ 0.59\% & 0.93 $\pm$ 0.31\% & 0.23 $\pm$ 0.03\% & 0.57 $\pm$ 0.40\% & 0.43 $\pm$ 0.01\% & 0.57 $\pm$ 0.14\%\\
ALMM & 2.47 $\pm$ 0.19\% & 0.88 $\pm$ 0.61\% & 0.70 $\pm$ 0.22\% & 0.24 $\pm$ 0.02\% & 0.37 $\pm$ 0.29\% & 0.12 $\pm$ 0.02\% & 0.46 $\pm$ 0.12\%\\
\hline
DAEU & 3.27 $\pm$ 0.24\% & 0.49 $\pm$ 0.67\% & 0.67 $\pm$ 0.21\% & 0.25 $\pm$ 0.08\% & 0.30 $\pm$ 0.11\% & 0.40 $\pm$ 0.03\% & 0.42 $\pm$ 0.22\%\\
DAEN & 2.57 $\pm$ 0.11\% & 0.48 $\pm$ 0.24\% & 0.46 $\pm$ 0.27\% & 0.18 $\pm$ 0.05\% & 0.18 $\pm$ 0.34\% & 0.32 $\pm$ 0.01\% & 0.33 $\pm$ 0.09\%\\
CNNAEU & 2.49 $\pm$ 0.16\% & \bf 0.47 $\pm$ 0.51\% & 0.73 $\pm$ 0.26\% & 0.23 $\pm$ 0.07\% & 0.15 $\pm$ 0.13\% & 0.29 $\pm$ 0.01\% & 0.38 $\pm$ 0.27\%\\
\hline
EGU-Net-pw & 2.00 $\pm$ 0.10\% & 0.51 $\pm$ 0.32\% & 0.38 $\pm$ 0.19\% & 0.17 $\pm$ 0.01\% & \bf 0.13 $\pm$ 0.14\% & 0.40 $\pm$ 0.02\% & 0.32 $\pm$ 0.15\%\\
EGU-Net-ss & \bf 1.83 $\pm$ 0.07\% & 0.52 $\pm$ 0.27\% & \bf 0.33 $\pm$ 0.13\% & \bf 0.13 $\pm$ 0.04\% & 0.19 $\pm$ 0.18\% & \bf 0.20 $\pm$ 0.01\% & \bf 0.27 $\pm$ 0.12\%\\
\bottomrule[1.5pt]
\end{tabular}}
\label{table:unmixing_simulated}
\end{table*}

\begin{table*}
\centering
\caption{Quantitative results on the Jasper Ridge dataset, where aRMSE and aSAD as well as SAD for each material are reported with their standard deviations from 10 Monte Carlo runs. The best one is shown in bold.}
\resizebox{0.91\textwidth}{!}{ 
\begin{tabular}{l|c|cccc|c}
\toprule[1.5pt]
Methods & aRMSE & Tree & Water & Soil & Road & aSAD \\
\hline\hline
FCLSU & 0.1783 $\pm$ 0.0098 & 0.1580 $\pm$ 0.0336 & 0.1781 $\pm$ 0.0505 & 0.1179 $\pm$ 0.0171 & 0.0997 $\pm$ 0.0826 & 0.1384 $\pm$ 0.0177\\
PCLSU & 0.1673 $\pm$ 0.0083 & 0.1494 $\pm$ 0.0326 & 0.2236 $\pm$ 0.0695 & 0.1137 $\pm$ 0.0204 & 0.1327 $\pm$ 0.0418 & 0.1548 $\pm$ 0.0201\\
SUnSAL & 0.1649 $\pm$ 0.0069 & 0.1471 $\pm$ 0.0343 & 0.2374 $\pm$ 0.0717 & 0.1076 $\pm$ 0.0185 & 0.1199 $\pm$ 0.0348 & 0.1530 $\pm$ 0.0212\\
SULoRA & 0.1088 $\pm$ 0.0151 & 0.1242 $\pm$ 0.0360 & 0.2740 $\pm$ 0.1533 & 0.0786 $\pm$ 0.0345 & 0.1674 $\pm$ 0.0648 & 0.1611 $\pm$ 0.0532\\
ALMM & 0.1015 $\pm$ 0.0172 & 0.1091 $\pm$ 0.0401 & 0.2226 $\pm$ 0.0460 & 0.0837 $\pm$ 0.0148 & 0.1182 $\pm$ 0.0551 & 0.1334 $\pm$ 0.0174\\
\hline
DAEU & 0.1091 $\pm$ 0.0150 & 0.1532 $\pm$ 0.0316 & \bf 0.1245 $\pm$ 0.0810 & 0.1170 $\pm$ 0.0177 & 0.1077 $\pm$ 0.0232 & 0.1256 $\pm$ 0.0240\\
DAEN & 0.0957 $\pm$ 0.0234 & 0.1433 $\pm$ 0.0439 & 0.1305 $\pm$ 0.0317 & 0.0748 $\pm$ 0.0144 & 0.1005 $\pm$ 0.0483 & 0.1123 $\pm$ 0.0245\\
CNNAEU & 0.0934 $\pm$ 0.0211 & 0.1112 $\pm$ 0.0385 & 0.1544 $\pm$ 0.0913 & 0.0787 $\pm$ 0.0160 & \bf 0.1001 $\pm$ 0.0552 & 0.1111 $\pm$ 0.0367\\
\hline
EGU-Net-pw & 0.0896 $\pm$ 0.0226 & 0.1072 $\pm$ 0.0343 & 0.1733 $\pm$ 0.0748 & 0.0706 $\pm$ 0.0118 & 0.1052 $\pm$ 0.0540 & 0.1141 $\pm$ 0.0388\\
EGU-Net-ss & \bf 0.0861 $\pm$ 0.0211 & \bf 0.1002 $\pm$ 0.0338 & 0.1684 $\pm$ 0.0725 & \bf 0.0622 $\pm$ 0.0155 & 0.1032 $\pm$ 0.0540 & \bf 0.1085 $\pm$ 0.0362\\
\bottomrule[1.5pt]
\end{tabular}}
\label{table:unmixing_river}
\end{table*}

\begin{table*}
\centering
\caption{Quantitative results on the EnMAP Munich dataset, where aRMSE and aSAD as well as SAD for each material are reported with their standard deviations from 10 Monte Carlo runs. The best one is shown in bold.}
\resizebox{1\textwidth}{!}{ 
\begin{tabular}{l|c|ccccc|c}
\toprule[1.5pt]
Methods & aRMSE & Roof & Asphalt & Soil & Water & Vegetation & aSAD \\
\hline\hline
FCLSU & 33.15 $\pm$ 1.85\% & 50.73 $\pm$ 16.99\% & 43.17 $\pm$ 11.99\% & 17.25 $\pm$ 4.58\% & 12.97 $\pm$ 3.08\% & 17.63 $\pm$ 5.29\% & 28.35 $\pm$ 3.04\%\\
PCLSU & 24.73 $\pm$ 4.72\% & 34.63 $\pm$ 20.06\% & 28.27 $\pm$ 13.79\% & 15.23 $\pm$ 4.67\% & 32.69 $\pm$ 14.06\% & 12.81 $\pm$ 6.72\% & 24.73 $\pm$ 4.64\%\\
SUnSAL & 23.98 $\pm$ 4.06\% & 31.32 $\pm$ 20.67\% & 25.86 $\pm$ 13.56\% & 13.93 $\pm$ 3.62\% & 35.89 $\pm$ 14.04\% & 11.73 $\pm$ 6.94\% & 23.74 $\pm$ 5.18\%\\
SULoRA & 22.48 $\pm$ 4.74\% & 19.32 $\pm$ 10.21\% & 24.31 $\pm$ 23.88\% & 12.57 $\pm$ 3.80\% & 21.93 $\pm$ 13.02\% & 9.48 $\pm$ 10.12\% & 17.52 $\pm$ 3.23\%\\
ALMM & 22.97 $\pm$ 3.15\% & 18.95 $\pm$ 10.98\% & 14.32 $\pm$ 6.83\% & 15.32 $\pm$ 0.42\% & 13.73 $\pm$ 0.79\% & 13.13 $\pm$ 7.11\% & 15.09 $\pm$ 3.36\%\\
\hline
DAEU & 23.55 $\pm$ 4.33\% & 29.23 $\pm$ 21.81\% & 28.38 $\pm$ 19.88\% & 9.56 $\pm$ 2.70\% & 15.68 $\pm$ 7.39\% & 13.89 $\pm$ 9.63\% & 19.35 $\pm$ 6.39\%\\
DAEN & 21.18 $\pm$ 4.36\% & 19.37 $\pm$ 8.45\% & 16.54 $\pm$ 9.27\% & 15.33 $\pm$ 7.03\% & 20.75 $\pm$ 6.01\% & 10.68 $\pm$ 15.72\% & 16.53 $\pm$ 4.37\%\\
CNNAEU & 18.34 $\pm$ 1.34\% & 6.22 $\pm$ 1.93\% & 4.53 $\pm$ 2.00\% & 6.36 $\pm$ 0.66\% & 10.64 $\pm$ 1.51\% & 4.01 $\pm$ 2.32\% & 6.35 $\pm$ 1.25\%\\
\hline
EGU-Net-pw & 17.49 $\pm$ 1.40\% & \bf 5.60 $\pm$ 1.32\% & 4.13 $\pm$ 2.40\% & 6.28 $\pm$ 0.40\% & 10.00 $\pm$ 0.56\% & \bf 3.26 $\pm$ 1.38\% & \bf 5.85 $\pm$ 1.52\%\\
EGU-Net-ss & \bf 16.80 $\pm$ 1.07\% & 6.02 $\pm$ 0.90\% & \bf 3.88 $\pm$ 1.38\% & \bf 6.08 $\pm$ 0.42\% & \bf 9.08 $\pm$ 0.31\% & 5.42 $\pm$ 1.43\% & 6.13 $\pm$ 1.12\%\\
\bottomrule[1.5pt]
\end{tabular}}
\label{table:unmixing_munich}
\end{table*}

\begin{figure*}[!t]
	  \centering
		\subfigure{
			\includegraphics[width=0.99\textwidth]{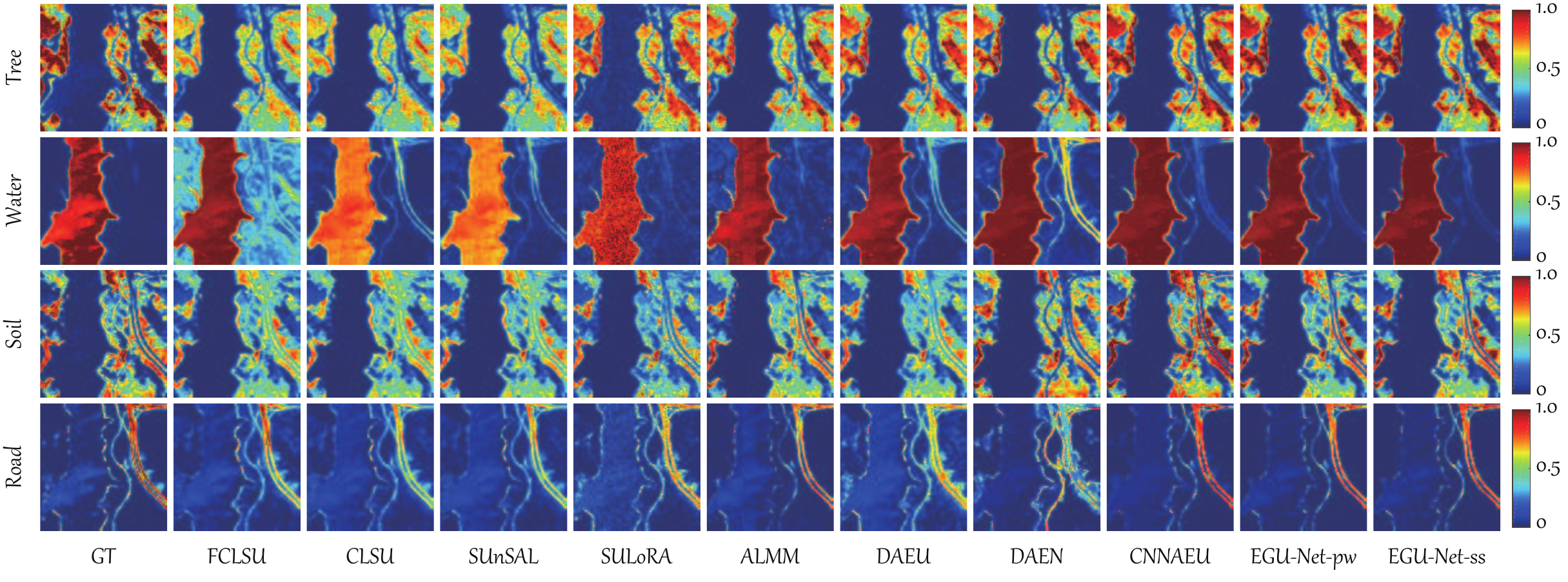}
		}
        \caption{Visualization of abundance maps using different unmixing methods on the AVIRIS Jasper Ridge data}
\label{fig:abundances_river}
\end{figure*}

\begin{figure*}[!t]
	  \centering
		\subfigure{
			\includegraphics[width=0.99\textwidth]{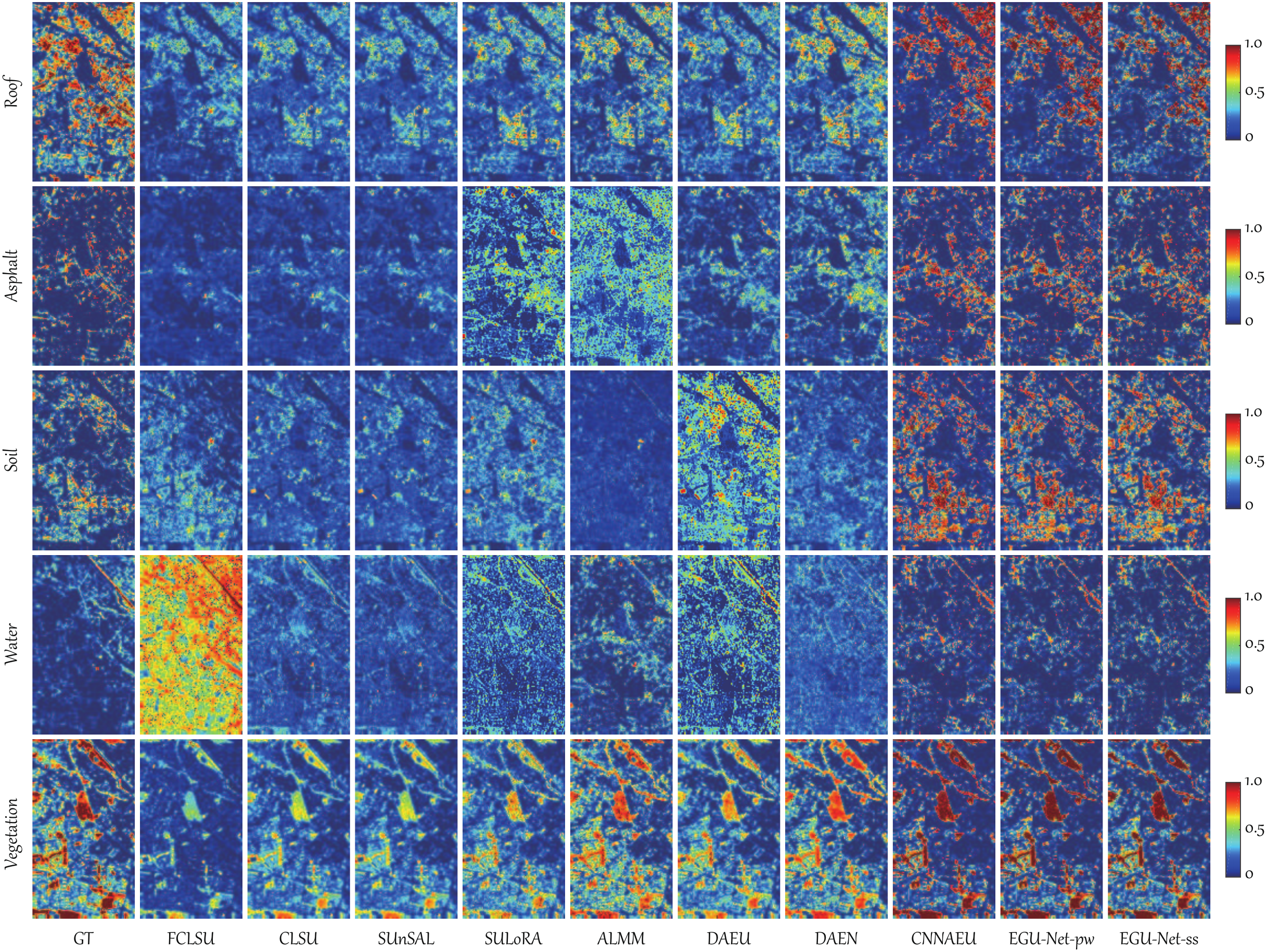}
		}
        \caption{Visualization of abundance maps using different unmixing methods on the EnMAP Munich data}
\label{fig:abundances_munich}
\end{figure*}

\begin{figure*}[!t]
	  \centering
		\subfigure[endmembers of AVIRIS Jasper Ridge data]{
			\includegraphics[width=0.38\textwidth]{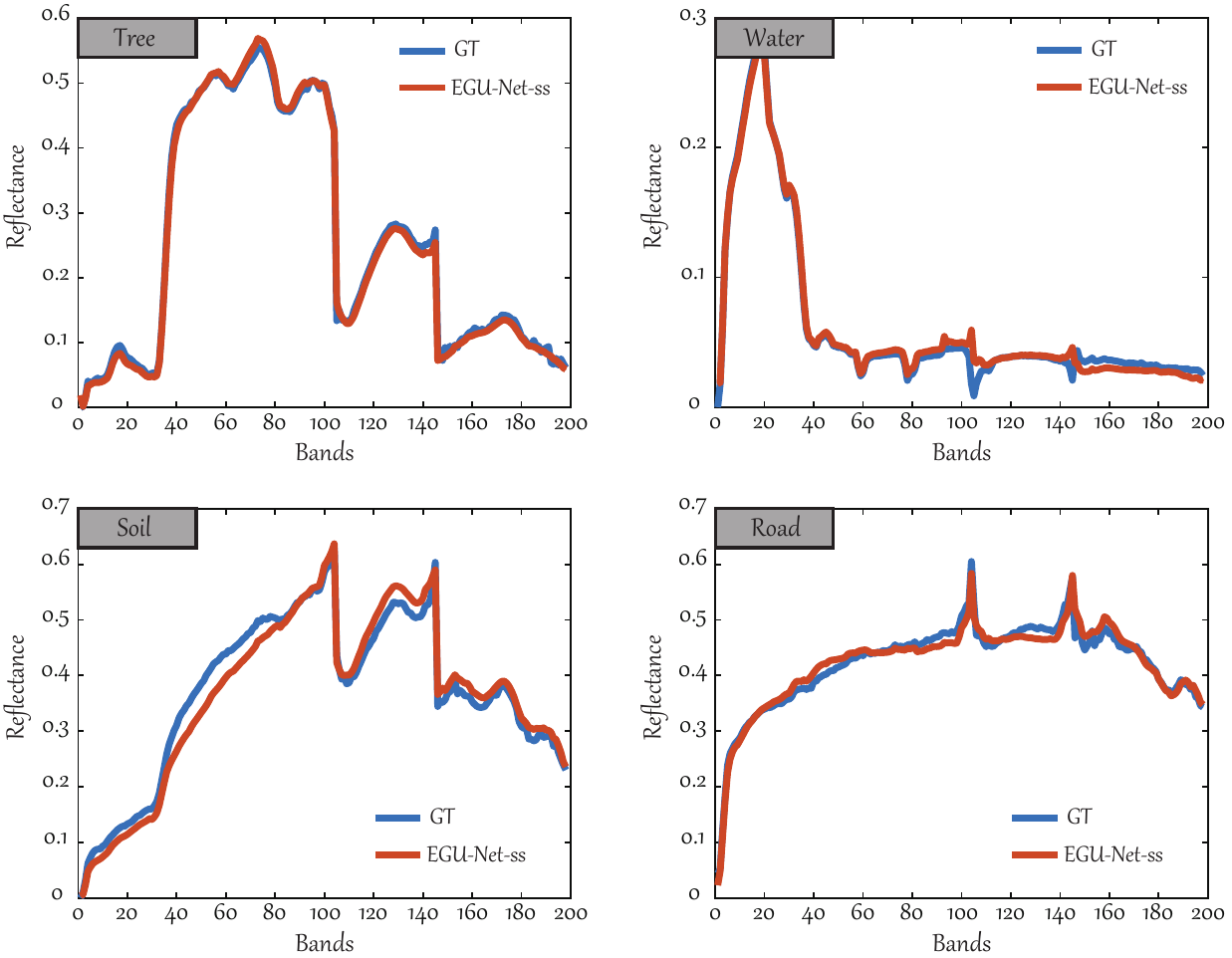}
			\label{fig:endmembers_river}
		}
		\subfigure[endmembers of EnMAP Munich data]{
			\includegraphics[width=0.57\textwidth]{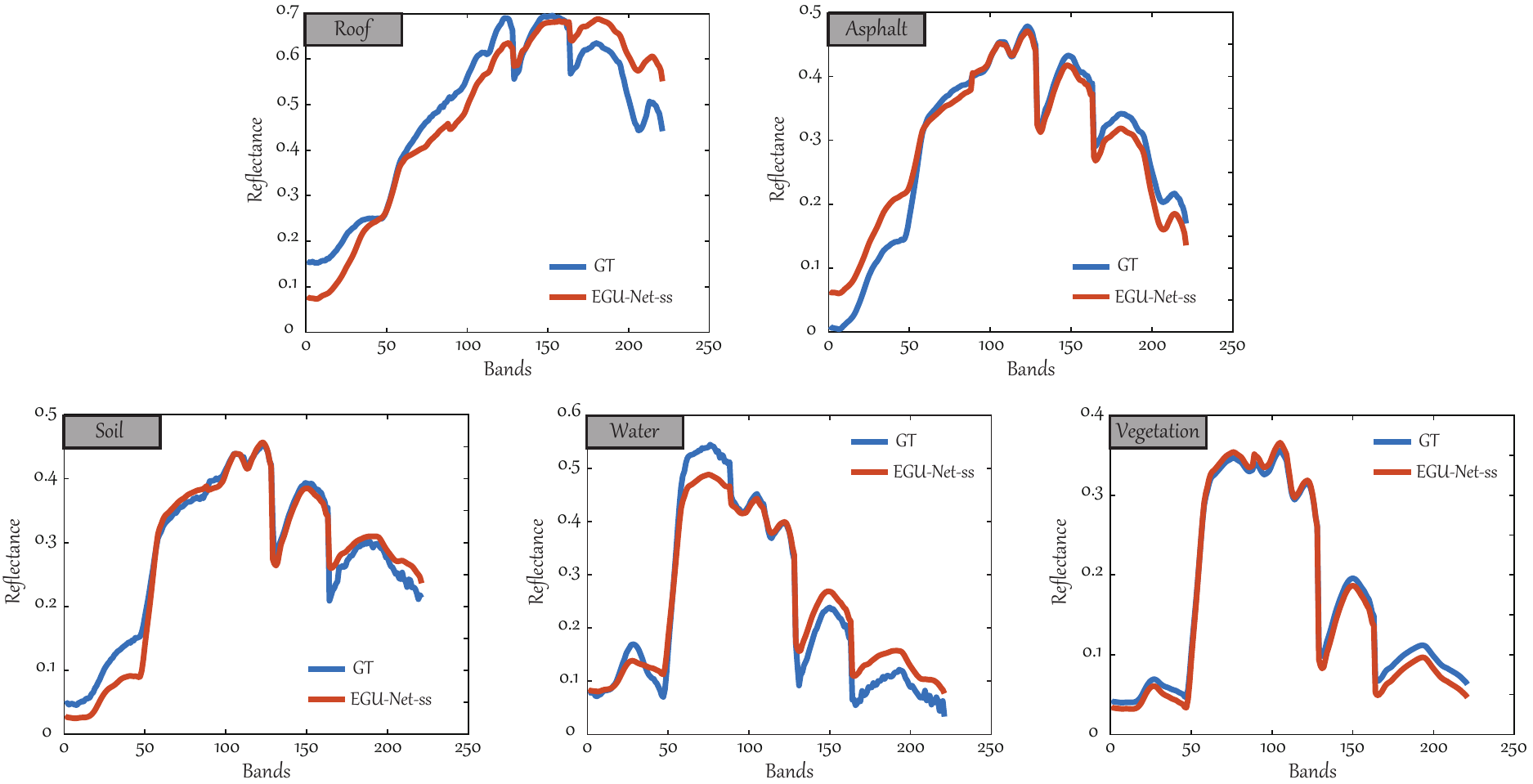}
			\label{fig:endmembers_munich}
		}
         \caption{Extracted endmember comparison between the proposed EGU-Net-ss and corresponding GT. (a) AVIRIS Jasper Ridge Data. (b) EnMAP Munich data.}
\label{fig:endmembers}
\end{figure*}

\subsection{Results and Analysis on the Real Data}

To further verify the effectiveness and generalization ability, we evaluate our network architecture on two more challenging real datasets: Jasper Ridge and Munich in comparison with the same compared methods. Similarly, the three indices: aRMSE, SAD for each endmember and their mean value aSAD are calculated for the quantitative assessment of unmixing results, as detailed in Tables \ref{table:unmixing_river} and \ref{table:unmixing_munich}.

By strictly following the ASC, FCLSU performs extremely poor estimation for the abundances and endmembers as well. One possible reason is that more complex SVs lie in the real HSIs. Although PCLSU outperforms FCLSU with an improvement of more than 0.01 and 9\% aRMSEs on the two datasets, respectively, owing to the effective relaxation of the ASC, it still fails to address the SVs. The sparsity-prompting constraint (SUnSAL) can not also play a big role in the blind HU of real scene. By designedly modeling the SVs, SULoRA and ALMM bring increments of at least 0.05 and 0.06 aRMSEs on the first real data as well as 1.5\% and 1\% aRMSEs on the second real data over SUnSAL.

For the DL-based unmixing networks, they are, by and large, superior to those traditional LMM-based methods either in aRMSE or SADs and aSAD on both datasets. In particular, this trend becomes more obvious in the real data. Remarkably, our EGU-Net overcomes other competitors from the perspective of either abundance estimation or endmember extraction at a further improvement of nearly 0.005 and 2.5\% aRMSEs on the basis of CNNAEU for the two real HSIs, respectively. In addition, there is a similar trend between-in EGU-Net-pw and EGU-Net-ss, that is, the latter that emphasizes simultaneous spatial-spectral modeling observably exceeds the former with a focus on pixel-wise strategy. This not only demonstrates the effectiveness and superiority of the proposed EGU-Net framework but also the use of convolutional operations (e.g., CNNs) in the blind HU tasks.

\subsection{Visual Evaluation}

We also perform a visual comparison of compared methods on the estimations of abundance maps and endmembers.
\subsubsection{Abundance Visualization} As shown in Fig. \ref{fig:abundances_river}, FCLSU fails to unmix the materials well, especially \textit{Water} leading to the large-area errors. By relaxing the strong ASC, the PCLSU's abundance map of \textit{Water} looks much better than FCLSU's, but the fractional values between PCLSU and ground truth (GT) still exist the large gap. The similar results also occurs in SUnSAL. By attempts to eliminate the effects of SVs, the abundance maps of SULoRA more approach to the GTs visually, while ALMM shows relatively smooth maps, particularly the material \textit{Water}. For those network-based unmixing models, they all behave more realistic on the abundance maps, although DAEU and DAEN wrongly recognize the \textit{Road} as \textit{Water} to some extent. As expected, our EGU-Net makes the basically identical visual effects with GTs, and meanwhile the EGU-Net-ss shows more robust and smooth maps. Furthermore, a similar trend can be observed in Fig. \ref{fig:abundances_munich} for the EnMAP Munich data.
This more complex urban scene lies in greater challenges. This, therefore, makes more noisy and easily-confused abundance maps. Despite so, our proposed networks are still able to visualize the nearly-identical shapes for each material compared to the GT.

\subsubsection{Endmember Comparison} With desirable abundance maps, we are also expected to have good endmember estimations, e.g., in shapes or band absorption. Towards this goal, we make a visual comparison for endmembers between our EGU-Net-ss and GTs, as shown in Fig. \ref{fig:endmembers}. The AVIRIS Jasper Ridge image is a easy-understanding scene, which only contains four common materials: \textit{Tree}, \textit{Water}, \textit{Soil}, and \textit{Road}. As a result, EGU-Net-ss performs relatively accurate endmember estimations (see Fig. \ref{fig:endmembers_river}). For another more challenging EnMAP Munich data, there exists a gap between the extracted endmembers and GTs. This indicates, to some extent, that the studied urban scene is complex, possibly involving many unknown SVs. We have to admit, however, that in such complex urban scene, our network is still capable of learning high-quality endmembers that can be matched well in shapes, e.g., \textit{Asphalt} and \textit{Vegetation}, as shown in Fig. \ref{fig:endmembers_munich}.

\section{Conclusion}
In this paper, we propose a general self-supervised unmixing network, called EGU-Net, which consists of EGU-Net-pw and EGU-Net-ss. The two resulting networks are applicable to pixel-wise spectral unmixing and spatial-spectral unmixing with fully connected encoder and convolutional operations, respectively. In addition, unlike the widely-used autoencoder-like models, EGU-Net additionally learns an \textit{E-Net} from pure or nearly-pure endmembers and transfers its parameters into the another autoencoder-based unmixing network: \textit{UR-Net}. In the meanwhile, the unmixing results of \textit{UR-Net} can be also fed back to the \textit{E-Net} to guide a better endmember learning, thereby yielding more reasonable and superior unmixing. In our future work, we would like to optimize the network architecture by using more advanced reconstruction loss (e.g., SAD or sparsity-promoting $\ell_{1}$-norm) and investigate the optimal combination of endmember extraction and unmixing methods used in \textit{E-Net} as well as develop a more general and powerful network-based framework with the aid of the multi-modal data (e.g., multispectral data, Lidar) to address this unmixing issue more effectively. 


\bibliographystyle{ieeetr}
\bibliography{HDF_ref}

\begin{IEEEbiography}[{\includegraphics[width=1in,height=1.25in,clip,keepaspectratio]{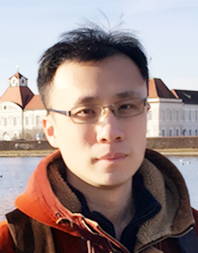}}]{Danfeng Hong}
(S'16--M'19--SM'21) received the M.Sc. degree (summa cum laude) in computer vision, College of Information Engineering, Qingdao University, Qingdao, China, in 2015, the Dr. -Ing degree (summa cum laude) in Signal Processing in Earth Observation (SiPEO), Technical University of Munich (TUM), Munich, Germany, in 2019. 

Since 2015, he has worked as a Research Associate at the Remote Sensing Technology Institute (IMF), German Aerospace Center (DLR), Oberpfaffenhofen, Germany. Currently, he is a research scientist and leads a Spectral Vision working group at IMF, DLR, and also an adjunct scientist in GIPSA-lab, Grenoble INP, CNRS, Univ. Grenoble Alpes, Grenoble, France. His research interests include signal / image processing and analysis, hyperspectral remote sensing, machine / deep learning, artificial intelligence and their applications in Earth Vision.

Dr. Hong was a recipient of the Best Reviewer award of the IEEE TRANSACTIONS ON GEOSCIENCE AND REMOTE SENSING in 2020, and the Jose Bioucas Dias award for recognizing the outstanding paper at the Workshop on Hyperspectral Imaging and Signal Processing: Evolution in Remote Sensing (WHISPERS) in 2021. He is a Guest Editor for the IEEE JOURNAL OF SELECTED TOPICS IN APPLIED EARTH OBSERVATIONS AND REMOTE SENSING and REMOTE SENSING. Currently, he is a Editorial aboard member of the REMOTE SENSING and a Topical Associate Editor of the IEEE TRANSACTIONS ON GEOSCIENCE AND REMOTE SENSING.
\end{IEEEbiography}

\vskip -2\baselineskip plus -1fil

\begin{IEEEbiography}[{\includegraphics[width=1in,height=1.25in,clip,keepaspectratio]{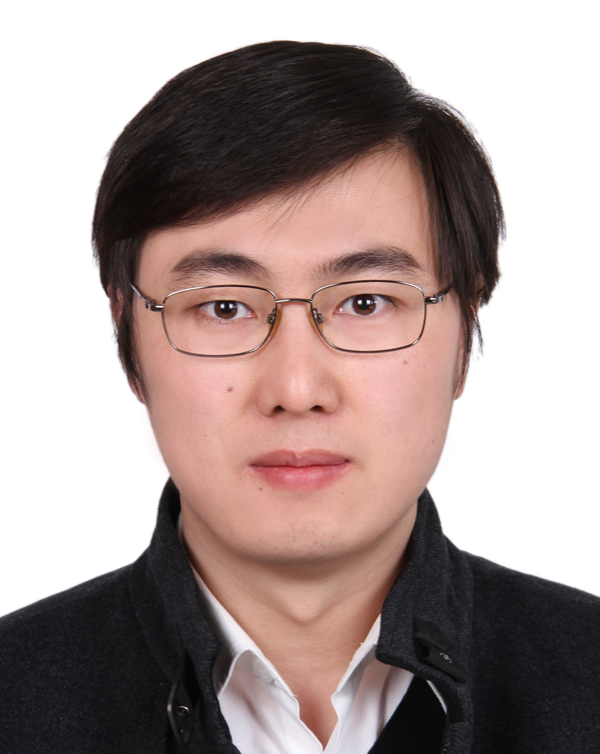}}]{Lianru Gao} (M'12--SM'18) received the B.S. degree in civil engineering from Tsinghua University, Beijing, China, in 2002, the Ph.D. degree in cartography and geographic information system from Institute of Remote Sensing Applications, Chinese Academy of Sciences (CAS), Beijing, China, in 2007.

He is currently a Professor with the Key Laboratory of Digital Earth Science, Aerospace Information Research Institute, CAS. He also has been a visiting scholar at the University of Extremadura, Cáceres, Spain, in 2014, and at the Mississippi State University (MSU), Starkville, USA, in 2016. His research focuses on hyperspectral image processing and information extraction. In last ten years, he was the PI of 10 scientific research projects at national and ministerial levels, including projects by the National Natural Science Foundation of China (2010-2012, 2016-2019, 2018-2020), and by the Key Research Program of the CAS (2013-2015). He has published more than 160 peer-reviewed papers, and there are more than 80 journal papers included by SCI. He was coauthor of an academic book entitled ``Hyperspectral Image Classification And Target Detection''. He obtained 28 National Invention Patents in China. He was awarded the Outstanding Science and Technology Achievement Prize of the CAS in 2016, and was supported by the China National Science Fund for Excellent Young Scholars in 2017, and won the Second Prize of The State Scientific and Technological Progress Award in 2018. He received the recognition of the Best Reviewers of the IEEE Journal of Selected Topics in Applied Earth Observations and Remote Sensing in 2015, and the Best Reviewers of the IEEE Transactions on Geoscience and Remote Sensing in 2017.
\end{IEEEbiography}

\vskip -2\baselineskip plus -1fil

\begin{IEEEbiography}[{\includegraphics[width=1in,height=1.25in,clip,keepaspectratio]{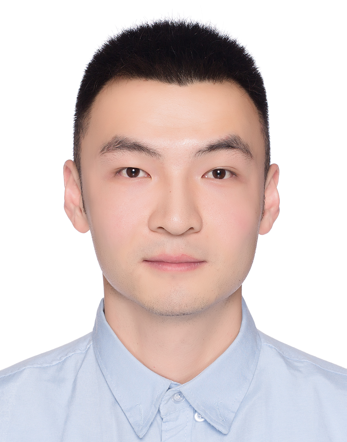}}]{Jing Yao} received the B.Sc. degree from Northwest University, Xi’an, China, in 2014, and the Ph.D. degree in the School of Mathematics and Statistics, Xi’an Jiaotong University, Xi’an, China, in 2021. 

He is currently an Assistant Professor with the Key Laboratory of Digital Earth Science, Aerospace Information Research Institute, Chinese Academy of Sciences, Beijing, China. From 2019 to 2020, he was a visiting student at Signal Processing in Earth Observation (SiPEO), Technical University of Munich (TUM), Munich, Germany, and at the Remote Sensing Technology Institute (IMF), German Aerospace Center (DLR), Oberpfaffenhofen, Germany.

His research interests include low-rank modeling, hyperspectral image analysis and deep learning-based image processing methods.
\end{IEEEbiography}

\vskip -2\baselineskip plus -1fil

\begin{IEEEbiography}[{\includegraphics[width=1in,height=1.25in,clip,keepaspectratio]{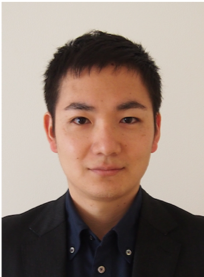}}]{Naoto Yokoya} (S'10--M'13) received the M.Eng. and Ph.D. degrees from the Department of Aeronautics and Astronautics, the University of Tokyo, Tokyo, Japan, in 2010 and 2013, respectively.

He is currently a Lecturer at the University of Tokyo and a Unit Leader at the RIKEN Center for Advanced Intelligence Project, Tokyo, Japan, where he leads the Geoinformatics Unit. He was an Assistant Professor at the University of Tokyo from 2013 to 2017. In 2015-2017, he was an Alexander von Humboldt Fellow, working at the German Aerospace Center (DLR), Oberpfaffenhofen, and Technical University of Munich (TUM), Munich, Germany. His research is focused on the development of image processing, data fusion, and machine learning algorithms for understanding remote sensing images, with applications to disaster management.

Dr. Yokoya won the first place in the 2017 IEEE Geoscience and Remote Sensing Society (GRSS) Data Fusion Contest organized by the Image Analysis and Data Fusion Technical Committee (IADF TC). He is the Chair (2019-2021) and was a Co-Chair (2017-2019) of IEEE GRSS IADF TC and also the secretary of the IEEE GRSS All Japan Joint Chapter since 2018. He is an Associate Editor for the IEEE Journal of Selected Topics in Applied Earth Observations and Remote Sensing (JSTARS) since 2018. He is/was a Guest Editor for the IEEE JSTARS in 2015-2016, for Remote Sensing in 2016-2020, and for the IEEE Geoscience and Remote Sensing Letters (GRSL) in 2018-2019.
\end{IEEEbiography}

\vskip -2\baselineskip plus -1fil

\begin{IEEEbiography}[{\includegraphics[width=1in,height=1.25in,clip,keepaspectratio]{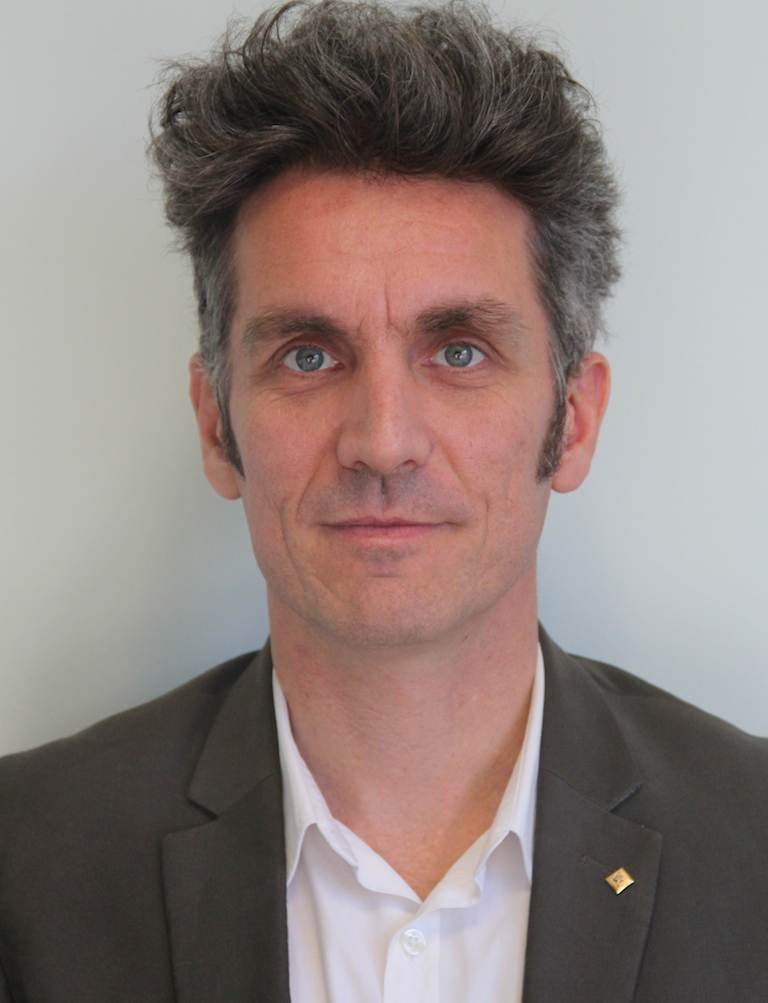}}]{Jocelyn Chanussot}
(M'04--SM'04--F'12) received the M.Sc. degree in electrical engineering from the Grenoble Institute of Technology (Grenoble INP), Grenoble, France, in 1995, and the Ph.D. degree from the Université de Savoie, Annecy, France, in 1998. Since 1999, he has been with Grenoble INP, where he is currently a Professor of signal and image processing. His research interests include image analysis, hyperspectral remote sensing, data fusion, machine learning and artificial intelligence. He has been a visiting scholar at Stanford University (USA), KTH (Sweden) and NUS (Singapore). Since 2013, he is an Adjunct Professor of the University of Iceland. In 2015-2017, he was a visiting professor at the University of California, Los Angeles (UCLA). He holds the AXA chair in remote sensing and is an Adjunct professor at the Chinese Academy of Sciences, Aerospace Information research Institute, Beijing.

Dr. Chanussot is the founding President of IEEE Geoscience and Remote Sensing French chapter (2007-2010) which received the 2010 IEEE GRS-S Chapter Excellence Award. He has received multiple outstanding paper awards. He was the Vice-President of the IEEE Geoscience and Remote Sensing Society, in charge of meetings and symposia (2017-2019). He was the General Chair of the first IEEE GRSS Workshop on Hyperspectral Image and Signal Processing, Evolution in Remote sensing (WHISPERS). He was the Chair (2009-2011) and  Cochair of the GRS Data Fusion Technical Committee (2005-2008). He was a member of the Machine Learning for Signal Processing Technical Committee of the IEEE Signal Processing Society (2006-2008) and the Program Chair of the IEEE International Workshop on Machine Learning for Signal Processing (2009). He is an Associate Editor for the IEEE Transactions on Geoscience and Remote Sensing, the IEEE Transactions on Image Processing and the Proceedings of the IEEE. He was the Editor-in-Chief of the IEEE Journal of Selected Topics in Applied Earth Observations and Remote Sensing (2011-2015). In 2014 he served as a Guest Editor for the IEEE Signal Processing Magazine. He is a Fellow of the IEEE, a member of the Institut Universitaire de France (2012-2017) and a Highly Cited Researcher (Clarivate Analytics/Thomson Reuters).
\end{IEEEbiography}

\vskip -2\baselineskip plus -1fil

\begin{IEEEbiography}[{\includegraphics[width=1in,height=1.25in,clip,keepaspectratio]{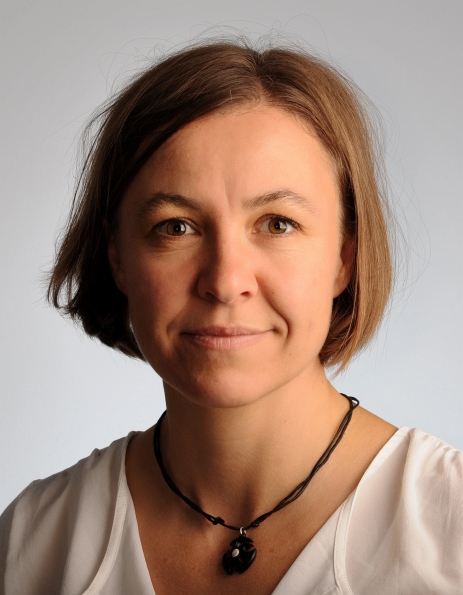}}]{Uta Heiden} (IEEE Member) received the Ph.D. degree in urban spectroscopy from the Technical University of Berlin, Berlin, Germany, and the GFZ German Research Centre for Geosciences, Potsdam, Germany, in 2003. She is also a Researcher with the Remote Sensing Technology Institute (IMF), German Aerospace Center (DLR), Weßling, Germany. She is also interested in application development using airborne and spaceborne imaging spectroscopy data with a focus on the use of material-based reflectance properties of urban surfaces. She has widened her interests toward a better understanding of the function of soils in degrading ecosystems. 

Dr. Heiden is also a member of the Science Advisory Group of the German spaceborne imaging spectrometer mission EnMAP. She is also a Science Coordinator for the spaceborne imaging spectroscopy mission DESIS onboard the ISS and the Chair of the ``Geoscience Spaceborne Imaging Spectroscopy'' Technical Committee (GSIS-TC) of the IEEE Geoscience and Remote Sensing Society.
\end{IEEEbiography}

\vskip -2\baselineskip plus -1fil

\begin{IEEEbiography}[{\includegraphics[width=1in,height=1.25in,clip,keepaspectratio]{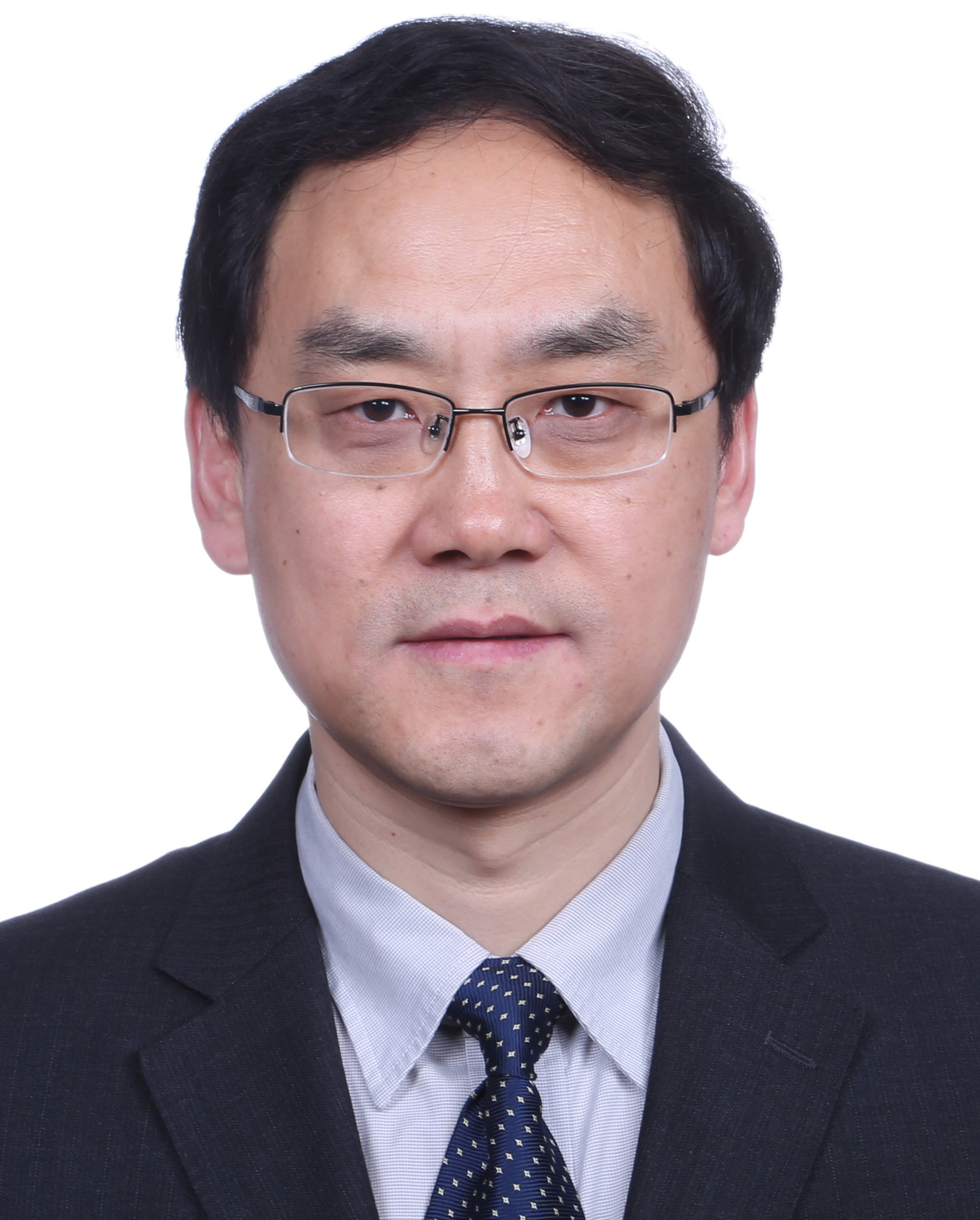}}]{Bing Zhang} (M'11--SM'12--F'19) received the B.S. degree in geography from Peking University, Beijing, China, in 1991, and the M.S. and Ph.D. degrees in remote sensing from the Institute of Remote Sensing Applications, Chinese Academy of Sciences (CAS), Beijing, China, in 1994 and 2003, respectively.

Currently, he is a Full Professor and the Deputy Director of the Aerospace Information Research Institute, CAS, where he has been leading lots of key scientific projects in the area of hyperspectral remote sensing for more than 25 years. His research interests include the development of Mathematical and Physical models and image processing software for the analysis of hyperspectral remote sensing data in many different areas. He has developed 5 software systems in the image processing and applications. His creative achievements were rewarded 10 important prizes from Chinese government, and special government allowances of the Chinese State Council. He was awarded the National Science Foundation for Distinguished Young Scholars of China in 2013, and was awarded the 2016 Outstanding Science and Technology Achievement Prize of the Chinese Academy of Sciences, the highest level of Awards for the CAS scholars.

Dr. Zhang has authored more than 300 publications, including more than 170 journal papers. He has edited 6 books/contributed book chapters on hyperspectral image processing and subsequent applications. He is the IEEE fellow and currently serving as the Associate Editor for IEEE Journal of Selected Topics in Applied Earth Observations and Remote Sensing. He has been serving as Technical Committee Member of IEEE Workshop on Hyperspectral Image and Signal Processing since 2011, and as the president of hyperspectral remote sensing committee of China National Committee of International Society for Digital Earth since 2012, and as the Standing Director of Chinese Society of Space Research (CSSR) since 2016. He is the Student Paper Competition Committee member in IGARSS from 2015-2019.
\end{IEEEbiography}

\end{document}